\documentclass[10pt]{article}
\pdfoutput=1

\usepackage{mathy}
\usepackage{amsmath,amssymb,amsthm,latexsym,bbold,calc,wasysym,mathtools,empheq,upgreek}
\usepackage{pifont,dsfont,marvosym,sidecap,stmaryrd,enumitem,nicefrac}
\usepackage[english]{babel}
\usepackage[usenames,dvipsnames]{xcolor}
\usepackage{relsize,graphicx,array,tabularx}
\usepackage[export]{adjustbox}
\usepackage[ampersand]{easylist}

\usepackage{tikz}
\usepackage{tikz-cd}
\usetikzlibrary{shapes.geometric}
\usetikzlibrary{arrows,matrix,calc,scopes,decorations.markings,intersections}
\usetikzlibrary{arrows.meta} 	
\usetikzlibrary{decorations.pathreplacing,angles,quotes,patterns,snakes}

\usepackage{amsthm}
\newtheoremstyle{mydef}%
	{0.9em} 
	{0.7em}
	{\itshape\hangindent=1.6em}
	{1.5em}
	{\scshape}
	{.}
	{.5em}
	{}%
	
\theoremstyle{mydef} 
\newtheorem{definition}{Definition}
\numberwithin{definition}{section}

\theoremstyle{mydef}

\numberwithin{lemma}{section}

\theoremstyle{mydef}

\numberwithin{theorem}{section}

\theoremstyle{mydef}
\newtheorem{example}{Example}
\numberwithin{example}{section}

\theoremstyle{mydef}

\numberwithin{proposition}{section}

\newcommand{\includeTikz}[3]{
	\includegraphics[scale=1, valign=c, raise=#1 pt]{fig/#2}
}

\newcommand{\rU}{{\rm U}}
\newcommand{\la}{\langle}
\newcommand{\ra}{\rangle}
\newcommand{\q}{\quad}

\newcommand{\sss}{\scriptstyle}
\newcommand{\ssss}{\scriptscriptstyle}
\newcommand{\bul}{\raisebox{1pt}{${\ssss \bullet}$}}
\newcommand{\snum}[1]{\text{\small $#1$}}
\newcommand{\ftnum}[1]{\text{\footnotesize $#1$}}
\newcommand{\mc}[1]{\mathcal{#1}}
\newcommand{\fr}[1]{\mathfrak{#1}}
\newcommand{\msf}[1]{\mathsf{#1}}
\newcommand{\pentsss}{\scalebox{0.88}{${\pentagon}$}}
\newcommand{\pentssss}{\scalebox{0.76}{${\pentagon}$}}
\DeclareMathOperator{\Fun}{\mathsf{Fun}}
\DeclareMathOperator{\Ob}{{\rm Ob}}
\DeclareMathOperator{\sOb}{{\rm sOb}}
\DeclareMathOperator{\Hom}{{\rm Hom}}
\DeclareMathOperator{\HomC}{\msf{Hom}}
\newcommand{\act}{\triangleright}
\newcommand{\cat}{\triangleleft}
\newcommand{\caact}{\bowtie}
\DeclareMathOperator{\actb}{\mathrlap{\hspace{0.45pt}\cdot}\triangleright}
\DeclareMathOperator{\catb}{\mathrlap{\hspace{1.6pt}\cdot}\triangleleft}
\DeclareMathOperator{\catbsss}{\mathrlap{\hspace{1.3pt}\cdot}\triangleleft}
\DeclareMathOperator{\Mod}{\mathsf{Mod}}
\DeclareMathOperator{\MOD}{\mathsf{MOD}}
\DeclareMathOperator{\Vect}{\mathsf{Vec}}
\DeclareMathOperator{\TVect}{\mathsf{2Vec}}
\DeclareMathOperator{\Rep}{\mathsf{Rep}}
\DeclareMathOperator{\TRep}{\mathsf{2Rep}}
\newcommand{\SixJAct}[6]{\big\{\!\begin{smallmatrix}#1 & #2 & #3\\#4 & #5 & #6 \end{smallmatrix}\!\big\}^{\act}} 
\newcommand{\SixJCat}[6]{\big\{\!\begin{smallmatrix}#1 & #2 & #3\\#4 & #5 & #6 \end{smallmatrix}\!\big\}^{\cat}} 
\newcommand{\SixJCaact}[6]{\big\{\!\begin{smallmatrix}#1 & #2 & #3\\#4 & #5 & #6 \end{smallmatrix}\!\big\}^{\caact}} 
\newcommand{\SixJ}[6]{\big\{\!\begin{smallmatrix}#1 & #2 & #3\\#4 & #5 & #6 \end{smallmatrix}\!\big\}} 
\newcommand{\VG}{\TVect_G^\pi}

\makeatletter
\newcommand{\xRrightarrow}[2][]{\ext@arrow 0359\Rrightarrowfill@{#1}{#2}}
\newcommand{\Rrightarrowfill@}{\arrowfill@\equiv\equiv\Rrightarrow}
\newcommand{\xLleftarrow}[2][]{\ext@arrow 3095\Lleftarrowfill@{#1}{#2}}
\newcommand{\Lleftarrowfill@}{\arrowfill@\Lleftarrow\equiv\equiv}
\makeatother

\usepackage{soul}

\hypersetup{%
	unicode=true,           
	pdftoolbar=true,        
	pdfmenubar=true,        
	pdffitwindow=false,     
	pdfstartview={FitH},    
	pdftitle={Tensor network approach to electromagnetic duality in (3+1)d topological gauge models},
	pdfauthor={Clement Delcamp},
	pdfsubject={},          
	pdfcreator={},          
	pdfproducer={pdfLaTeX}, 
	pdfkeywords={topological phases} {category theory} {tensor networks} {toric code} {duality} {module 2-category},
	pdfnewwindow=true,      
	colorlinks,
	citecolor=BrickRed,
	filecolor=BrickRed,
	linkcolor=BlueViolet,
	urlcolor=BrickRed,
	linktocpage=true
}

\title{\boldmath Tensor network approach to electromagnetic duality \par in (3+1)d topological gauge models}

\author[\pentagon, \hexagon]{Clement Delcamp}

\affiliation[\normalfont \pentagon]{Max Planck Institute of Quantum Optics \\ Hans-Kopfermann-Stra{\ss}e 1, 85748 Garching, Germany}
\affiliation[\normalfont \hexagon]{Max Planck Institute for the Physics of Complex Systems\\ Nöthnitzer Stra{\ss}e 38, 01187 Dresden, Germany}

\emailAdd{delcamp@pks.mpg.de}

\abstract{\\~\\  
	Given the Hamiltonian realisation of a topological (3+1)d gauge theory with finite group $G$, we consider a family of tensor network representations of its ground state subspace. This family is  indexed by gapped boundary conditions encoded into module 2-categories over the input spherical fusion 2-category.  Individual tensors are characterised by symmetry conditions with respect to non-local operators acting on entanglement degrees of freedom. 
	In the case of Dirichlet and Neumann boundary conditions, we show that the symmetry operators form the  fusion 2-categories $\TVect_G$ of $G$-graded 2-vector spaces and $\TRep(G)$ of 2-representations of $G$, respectively.
	In virtue of the Morita equivalence between $\TVect_G$ and $\TRep(G)$---which we explicitly establish---the topological order can be realised as the Drinfel'd centre of either 2-category of  operators; this is a realisation of the electromagnetic duality of the theory.
	Specialising to the case $G = \mathbb Z_2$, we recover tensor network representations that were recently introduced, as well as the relation between the electromagnetic duality of a pure (3+1)d $\mathbb Z_2$ gauge theory and the Kramers-Wannier duality of a boundary (2+1)d Ising model.}

\notoc

\begin{document} 
	\vspace*{-2em}
	\maketitle
	\flushbottom
	\newpage
	
\section{Introduction}

The three-dimensional \emph{toric code}  \cite{Kitaev1997,Hamma:2004ud} is arguably the simplest physical system in three spatial dimensions exhibiting \emph{topological order} \cite{Wen:1989iv,Chen:2010gda}. It can be conveniently defined as the Hamiltonian realisation of the (3+1)d \emph{Dijkgraaf-Witten} state-sum invariant with input group $\mathbb Z_2$ \cite{dijkgraaf1990topological} and its topological order is uniquely characterised by the existence of an emergent boson as the single non-trivial bulk point-like excitation \cite{Johnson-Freyd:2020twl}. In addition to these bosonic (electric) \emph{charges}, the (3+1)d toric code famously hosts bulk loop-like (magnetic) \emph{fluxes} that link non-trivially with them. 

Given an open manifold, it is well understood that the toric code Hamiltonian can be extended to the boundary while preserving the gap. More precisely, we distinguish two gapped boundary conditions, which are obtained by condensing either the point-like charges (Dirichlet) or the loop-like fluxes (Neumann). By tuning parameters of the quasi-two-dimensional system, we may encounter a phase transition between these two condensates, which can be checked to be in the universality class of the (2+1)d (quantum) \emph{Ising} model \cite{Ji:2019jhk,Kong_2020}. This second order phase transition is characterised by the spontaneous breaking of a global (0-form) $\mathbb Z_2$-symmetry whose charges correspond to the point-like charges of the bounding (3+1)d toric code. But the (2+1)d Ising model is famously dual to \emph{Wegner’s $\mathbb Z_2$-gauge theory} \cite{Wegner:1984qt,RevModPhys.51.659,Fisher2004,Zhao:2020vdn}, with respect to which the second order phase transition may be characterised by the spontaneous breaking of a (generalised) global 1-form $\mathbb Z_2$-symmetry \cite{gaiotto2015generalized,Ji:2019jhk,Kong_2020,Zhao:2020vdn}.  Charges of this 1-form $\mathbb Z_2$-symmetry then correspond to the loop-like fluxes of the bounding (3+1)d toric code. This means that the duality between the (2+1)d Ising model and Wegner's $\mathbb Z_2$-gauge theory can be understood in terms of the \emph{electromagnetic duality} of a bounding topological gauge theory \cite{Freed:2018cec,Kong_2020}.

The previous statements were recently revisited from a \emph{tensor network} viewpoint. At a basic level, tensor networks are collections of tensors that are contracted together following patterns dictated by graphs. An individual tensor carries both \emph{virtual} indices, along which the tensor is contracted to its neighbours, and \emph{physical} indices, uncontracted indices labelled by orthonormal basis states of some specified vector spaces. Physically, the resulting tensor network is regarded as a wave function in the tensor product of all its physical vector spaces, thought as the state space of a quantum system. Such tensor network states turn out to provide a very powerful class of variational wave functions for the study of strongly-correlated many-body systems \cite{PhysRevLett.69.2863,perez2006matrix,Verstraete:2004cf}. The merit of this approach stems from the fact that \emph{global} \emph{correlation} patterns can be built out of the contraction of entanglement degrees of freedom of \emph{local} tensors. This makes tensor network states especially suited for the study of models of topological phases of matter for which the \emph{order} is microscopically encoded into patterns of \emph{long-range entanglement} \cite{PhysRevB.84.165139,PhysRevB.83.035107,SCHUCH20102153,PhysRevLett.111.090501,BUERSCHAPER2014447,Sahinoglu:2014upb,BULTINCK2017183,Bultinck_2017,Williamson:2017uzx}. The topological order then manifests itself in the existence of so-called \emph{virtual symmetries}, i.e. symmetries with respect to operators acting solely along virtual indices. 

In this context, the electromagnetic duality of the (3+1)d toric code is reflected into the existence of two dual tensor network \emph{representations} of the ground state subspace \cite{Williamson:2020hxw,Delcamp:2020rds}. These two representations are obtained by initially imposing either family of stabiliser constraints and are characterised by virtual symmetries with respect to string- and membrane-like operators, respectively. The symmetry breaking patterns of the boundary theories are then recovered in the fixed point sectors of the corresponding \emph{transfer matrices} \cite{Delcamp:2020rds}. Interestingly, we can choose two different tensor network representations over distinct submanifolds via the introduction of an \emph{intertwining tensor network} that only possesses virtual indices. When interpreted as an operator on the quasi-two-dimensional system at the boundary, this intertwiner effectively performs the mapping from the (2+1)d Ising model to Wegner's $\mathbb Z_2$ gauge theory.

Mathematically, the existence of distinct tensor network representations is directly related to the notion of equivalence between topological lattice models. This equivalence has so far been mainly studied in (2+1)d, where (bosonic) topological orders are encoded into \emph{modular tensor categories} for which the simple objects correspond to (bulk) topological string-like operators \cite{Kitaev:2006lla, Levin:2004mi}. Crucially, the same category theoretical data serve as input for the \emph{Reshetikhin-Turaev} construction of 3d topological quantum field theories \cite{Reshetikhin:1990pr,Reshetikhin:1991tc}. It was recently shown that for a Reshetikhin-Turaev theory that admits non-trivial gapped boundaries, the corresponding topological order can be realised as the \emph{Drinfel'd centre} $\mc Z(\mc C)$ of the \emph{spherical fusion category} $\mc C$ of boundary operators for any choice of gapped boundary condition \cite{Freed:2020qfy}. Such Reshetikhin-Turaev theories, which are known as \emph{Turaev-Viro-Barrett-Westbury} theories \cite{Turaev:1992hq,Barrett:1993ab}, are \emph{fully-extended}, i.e. they capture locality all the way down to the point. Consequently, they admit state-sum descriptions induced from the data of the spherical fusion categories of boundary operators, which can in turn be exploited in order to build \emph{lattice Hamiltonian realisations}. Such Hamiltonian realisations are usually referred to as \emph{string-net models} in the physical literature \cite{Levin:2004mi,alex2011stringnet}. 

Any boundary condition yielding the same topological order, it follows that the corresponding string-net models are dual to one another. Mathematically, this is the statement that the input spherical fusion categories are \emph{Morita equivalent} \cite{etingof2016tensor}. Two fusion categories $\mc C$ and $\mc D$ are said to be Morita equivalent if there exists an (exact) \emph{module category} $\mc M$ over $\mc C$ such that $\mc D^{\rm op}$ is equivalent to the category $\mc C^\star_\mc M$ of $\mc C$-module functors from $\mc M$ to itself. A choice of (finite semi-simple) module category $\mc M$ specifies a choice of gapped boundary condition, while the simple objects in $\mc C^\star_\mc M$ correspond to the topological boundary operators \cite{kitaev2012models}. Similarly, given a string-net model, tensor network representations are labelled by (finite semi-simple) module categories over the input spherical fusion category such that virtual operators now correspond to objects in the Morita equivalent category of module endofunctors \cite{Lootens:2020mso}.\footnote{The underlying philosophy is analogous to that employed in order to obtain a combinatorial description of a $d$-dimensional topological quantum field theory given a choice of gapped boundary condition, whereby the partition function is obtained by summing over vacua of the compactified theory along the ($d$$-$1)-skeleton \cite{Bhardwaj:2016clt}.} Therefore, topological order in a (2+1)d string-net model can be realised as the Drinfel'd centre of the category of virtual operators for any choice of tensor network representation. 

\bigskip \noindent
As part of an ongoing effort to better our understanding of duality relations in (3+1)d topological models, the purpose of the present manuscript is to initiate the generalisation of the previous considerations to three spatial dimensions. Topological orders in (3+1)d are described by \emph{non-degenerate braided fusion 2-categories} such that objects and morphisms correspond to the membrane- and string-like operators, respectively \cite{Kong:2014qka,kong2015boundarybulk,lan2017classification,johnsonfreyd2020classification}. This dichotomy of operators can be naively traced back to the graphical calculus of monoidal 2-categories, whereby objects are depicted as membranes and 1-morphisms as strings at the junctions of membranes. Analogously to the lower-dimensional scenario, a given (3+1)d topological order can be realised as the Drinfel'd centre of the \emph{spherical fusion 2-category} of boundary operators for any choice of gapped boundary condition \cite{lan2017classification,Kong:2019brm,johnsonfreyd2020classification,bullivant2021crossing}. These spherical fusion 2-categories of boundary operators in turn serve as input data for concrete lattice realisations of the topological phase \cite{dijkgraaf1990topological,Crane:1993if,Crane:1994ji,Yetter:1993dh,mackaay2000finite,Walker:2011mda,Cui:2016bmd,douglas2018fusion}. Furthermore, they are expected fulfil an appropriate categorification of the hereinabove notion of Morita equivalence with respect to \emph{module 2-categories} that label the boundary conditions.\footnote{In \cite{Bullivant:2020xhy}, an alternative description of (2+1)d gapped boundaries for (3+1)d gauge models was provided in terms of \emph{pseudo-algebra objects} in the input fusion 2-categories. These two descriptions are expected to be equivalent in virtue of a categorification of the well-known correspondence between indecomposable module categories and category of modules over algebra objects \cite{ostrik2002module,etingof2016tensor}. See \cite{decoppet2021finite} for recent progress in this direction.}

We shall focus our study on electromagnetic dualities for generalisations of the (3+1)d toric code obtained as lattice Hamiltonian realisations of the (untwisted) Dijkgraaf-Witten theories \cite{dijkgraaf1990topological} for arbitrary finite groups. The main advantage of dealing with such gauge models is that we can safely gloss over some subtleties inherent to spherical fusion 2-categories, allowing to focus on the duality relations. Given a finite group $G$, we distinguish for such models two canonical gapped boundary conditions obtained by condensing either all the (bulk) point-like charges or all the (bulk) loop-like fluxes. It has been expected that the corresponding boundary operators form the fusion 2-category $\TVect_G$ of \emph{$G$-graded 2-vector spaces} and the fusion 2-category $\TRep(G)$ of 2-representations of $G$, respectively \cite{Kong_2020}. The topological order can thus be realised as the Drinfel'd centre of either one of these fusion 2-categories, which implies---as we shall confirm---that these are Morita equivalent. \emph{This is the electromagnetic duality of (3+1)d topological gauge models.} By identifying $\TVect_G$ as the input spherical fusion 2-category of a lattice model, we can label these two boundary conditions by the  module 2-categories $\TVect_G$ and $\TVect$, respectively. The Morita equivalence between $\TVect_G$ and $\TRep(G)$ thus amounts to the equivalence between $\TRep(G)$ and $(\TVect_G)^\star_{\TVect}$ as fusion 2-categories, where $(\TVect_G)^\star_{\TVect}$ now refers to the fusion 2-category of $\TVect_G$-module 2-endofunctors of $\TVect$. Additional boundary conditions are labelled by (finite semi-simple) $\TVect_G$-module 2-categories $\mc M$ and the corresponding boundary operators are expected to be encoded into $(\TVect_G)^\star_{\mc M}$. In this spirit, we argue that for every choice of (finite semi-simple) module 2-category $\mc M$ over the input spherical fusion 2-category $\TVect_G$, one can construct a tensor network representation such that the non-vanishing components of the tensors evaluate to the matrix elements of the \emph{module pentagonator} of $\mc M$. The virtual symmetry operators should then be labelled by simple objects and simple 1-morphisms in $(\TVect_G)^\star_\mc M$ so that the topological order can be realised as the Drinfel'd centre of the 2-category of virtual operators for any choice of tensor network representation. We study in detail the canonical representations labelled by $\mc M =\TVect_G$ and $\mc M =\TVect$, proving that the 2-category of virtual operators are indeed $\TVect_G$ and $\TRep(G)$, respectively.  Specialising to the case $G=\mathbb Z_2$, this provides the mathematical framework underpinning the results established in \cite{Delcamp:2020rds} pertaining to the electromagnetic duality of the (3+1)d toric code.

Although our study focuses on topological lattice gauge theories, the framework we employ can be adapted to accommodate more general models. Such a generalisation would be particularly insightful in light of the interplay between dualities of quantum physical systems in $d$ dimensions and dualities of topological lattice field theories in $d$+1 dimensions, generalising the relation between Kramers-Wannier duality of the Ising model and the electromagnetic duality of a bounding toric code. This interplay can be made very explicit in the case of one-dimensional models via the use of tensor network representations. Indeed, it is shown in \cite{bond} how a 1d quantum model can be defined given a tensor network representation of a string-net model and an implicit choice of algebra of operators. This is a generalisation of the \emph{anyonic chain} construction \cite{PhysRevLett.98.160409,PhysRevB.87.235120,PhysRevLett.101.050401,PhysRevLett.103.070401,ardonneAnyonChain,Buican:2017rxc}. Keeping the algebra of operators fixed, using another tensor network representation of the same string-net model then yields a dual model. We expect a similar situation in higher dimensions, whereby dual tensor network representations of a certain topological model are employed to explicitly construct dual 2d quantum models.

\begin{center}
    {\bf Organisation of the paper}
\end{center}
We begin in sec.~\ref{sec:preliminaries} by reviewing basic notions of module categories before categorifying them. After sketching the corresponding graphical calculus, we provide a definition of gauge models of topological phases in terms of equivalence classes of membrane-nets. In sec.~\ref{sec:TN} we construct families of tensor network representations  labelled by boundary conditions data for the gauge models and reveal the virtual symmetry operators associated with these representations. We also establish in this section the Morita equivalence between the fusion 2-categories of virtual operators associated with the two canonical tensor network representations. We conclude our analysis by specialising to the case of the (3+1)d toric code. 

\newpage

\section{Topological gauge models\label{sec:preliminaries}}
\emph{In this section we set the stage by introducing some  category theoretical concepts that are used throughout this manuscript and sketch the corresponding graphical calculus. Exploiting this graphical calculus, we then define topological lattice models in (3+1)d that have a lattice gauge theory interpretation.}

\subsection{Category theoretical preliminaries}
Let us begin by fixing some notations and conventions. Given a ($\mathbb C$-linear) \emph{category} $\mc C$, we notate the sets of objects and morphisms in $\mc C$ via $\Ob(\mc C)$ and $\Hom(\mc C)$, respectively. The set of morphisms (hom-set) between two objects $A,B \in \Ob(\mc C)$ is further denoted by $\Hom_\mc C(A,B)$. For each triple $A,B,C \in \Ob(\mc C)$, the composition rule is written as $\circ : \Hom_{\mc C}(A,B) \times \Hom_{\mc C}(B,C) \to \Hom_{\mc C}(A,C)$ and the identity morphism associated with any object $A$ is notated via ${\rm id}_A \in \Hom_{\mc C}(A,A)$. Henceforth, we assume that every category is $\mathbb C$-linear. 

We define a monoidal category as a quadruple $\mc C \equiv (\mc C, \otimes, \mathbb 1,\alpha)$ that consists of a category $\mc C$, a monoidal binary functor $\otimes: \mc C \times \mc C \to \mc C$, a distinguished unit object $\mathbb 1 \in \Ob(\mc C)$, and a natural isomorphism $\alpha: (\bul \otimes \bul) \otimes \bul \xrightarrow{\sim} \bul \otimes (\bul \otimes \bul)$ satisfying the `pentagon axiom' (see \cite{etingof2016tensor} for more detail). The natural isomorphism $\alpha$ shall be referred to as the monoidal \emph{associator} and its components read $\alpha_{A,B,C} : (A \otimes B) \otimes C \xrightarrow{\sim} A \otimes (B \otimes C)$ for all $A,B,C \in \Ob(\mc C)$. Without loss of generality, the unit isomorphisms shall be taken to be identities, thus identifying $\mathbb 1 \otimes A$ and $A \otimes \mathbb 1$ with $A \in \Ob(\mc C)$. Moreover, we denote by $(\mc C^{\rm op}, \otimes^{\rm op}, \mathbb 1, \alpha^{\rm op})$ the monoidal category \emph{opposite} to $\mc C$ such that $A \otimes^{\rm op} B := B \otimes A$ and $\alpha^{\rm op}_{A,B,C} = \alpha^{-1}_{C,B,A}$. 
Henceforth we implicitly assume that every monoidal category is \emph{finite} and \emph{semi-simple}. 

A prototypical example of a monoidal category is the category $\Vect$ of finite dimensional complex vector spaces and linear maps. This category has a single simple object, namely the one-dimensional vector space $\mathbb C$, and the monoidal functor is provided by the usual tensor product over $\mathbb C$. A more sophisticated example is obtained as a categorification of the notion of (twisted) group algebra over $\mathbb C$ by promoting the field $\mathbb C$ to $\Vect$:

\begin{example}[Group-graded vector spaces]
    Given a finite group $G$ and a normalised group 3-cocycle in $H^3(G,\rU(1))$, the monoidal category $\Vect_G^\alpha$ is defined as the category whose objects are $G$-graded vector spaces of the form $V = \bigoplus_{g \in G} V_g$ and morphisms are grading preserving linear maps. The monoidal functor is defined on homogeneous components as
    \begin{equation*}
        (V \otimes W)_g = \bigoplus_{x \in G}V_x \otimes V_{x^{-1}g} 
    \end{equation*}
    with unit $\mathbb 1$ such that $\mathbb 1_g = \delta_{g,\mathbb 1_G}\mathbb C$. The monoidal associator is provided by 
    \begin{equation*}
        \alpha_{U_g,V_h,W_k} = \alpha(g,h,k) \cdot (\mathring{\alpha}_{U,V,W})_{ghk} : (U_g \otimes V_h) \otimes W_k \xrightarrow{\sim} U_g \otimes (V_h \otimes W_k) \, ,
    \end{equation*}
    for all $g,h,k \in G$, where $\mathring{\alpha}_{U,V,W}$ refers  to the canonical isomorphism $(U \otimes V) \otimes W \xrightarrow{\sim} U \otimes (V \otimes W)$ in $\Vect$. The $|G|$-many simple objects are the one-dimensional vector spaces $\mathbb C_g$, for every $g \in G$, satisfying $\Hom_{\Vect_G}(\mathbb C_g, \mathbb C_h) = \delta_{g,h} \mathbb C$ and $\mathbb C_g \otimes \mathbb C_h \simeq \mathbb C_{gh}$.
\end{example}

\noindent
The study carried out in the next section largely relies on the notion of \emph{module category} over a monoidal category, and categorification thereof. We shall now briefly review the usual construction before categorifying it. We invite the reader to consult \cite{etingof2016tensor} for a more thorough exposition.
\begin{definition}[Left module category]
    Given a monoidal category $\mc C \equiv (\mc C, \otimes, \mathbb 1, \alpha)$, a left $\mc C$-module category is defined as a triple $(\mc M, \act, \alpha^\act)$ that consists of a category $\mc M$, a binary action functor $\act: \mc C \times \mc M \to \mc M$ and a natural isomorphism $\alpha^\act: (\bul \otimes \bul)\act \bul \xrightarrow{\sim} \bul \act (\bul \act \bul)$ satisfying a `pentagon axiom' that involves the monoidal associator $\alpha$. The natural isomorphism $\alpha^\act$ shall be referred to as the left module associator and its components read $\alpha^\act_{A,B,M}:(A \otimes B)\act M \xrightarrow{\sim} A \act (B \act M)$ for all $A,B \in \Ob(\mc C)$ and $M \in \Ob(\mc M)$.
\end{definition}

\noindent
Notice that every monoidal category $\mc C$ defines a (left) module category over itself such that the action functor is provided by the monoidal structure in $\mc C$. In the following, we shall refer to this $\mc C$-module category as the \emph{regular} $\mc C$-module category. A more interesting family of examples goes as follows:

\begin{example}[$\Vect_G$-module categories \cite{ostrik2002module,naidu2007categorical}\label{ex:VecGModule}] 
	Let $A \subset G$ be a subgroup of $G$ and $\phi$ a normalised group 2-cocycle in $H^2(A, \rU(1))$. We consider the category $\mc M$ whose collection of simple objects is the set $G/A$ of left cosets. Given the surjection $\msf p : G \to G / A$, $g \mapsto gA$, we denote by $\msf r : G / A \to G$ the map that assigns to every $M \in G /A$ a representative $\msf r(M) \in G$ such that $\msf p(\msf r(M))=M$.     Given the above, we define a binary functor $\act : \Vect_G \times \mc M \to \mc M$ via $\mathbb C_g \act M := \msf p(g \msf r(M))$, for all $g \in G$ and $M \in G /A$. In general, we have $g \msf r(M) \neq \msf r(\mathbb C_g \act M)$ and we notate via $a_{g,M}$ the group element in $A$ such that $g \msf r(M) =  \msf r (\mathbb C_g \act M) a_{g,M}$, for all $g \in G$ and  $M \in G /A$. It follows from the associativity of the group multiplication that
	\begin{equation}
		\label{eq:1cocycleConda}
		a_{gh,M} = a_{g,\mathbb C_h \act M} \, a_{h, M} \, , \q \forall \, g,h \in G \; {\rm and} \; M \in G /A \, .
	\end{equation}
	Noticing that the set of functions $\Hom(G/A, \mathbb C^\times)$ can be endowed with a (right) $G$-module structure via $(f \cat g)(M) := f(\mathbb C_g \act M)$, for any $g \in G$ and $M \in G/A$, we consider the 2-cochain $\alpha^{\act} \in C^2(G, \Hom(G/A,\mathbb C^\times))$ defined by $\alpha^{\act}(g,h)(M) := \phi(a_{g, \mathbb C_h \act M}, a_{h ,M})$ for any $g,h \in G$ and $\,M \in G/A$. It follows from eq.~\eqref{eq:1cocycleConda} satisfied by $a_{\bul,\bul}$ as well as the cocycle condition ${\rm d}\phi = 1$ that
	\begin{align}
		\alpha^{\act}(h,k)(M)\, \alpha^{\act}(g,hk)(M) = \alpha^{\act}(gh,k)(M) \, \alpha^{\act}(g,h)(\mathbb C_k \act M) \, ,
	\end{align}
	for any $g,h,k \in G$ and $M \in G/A$. i.e. $\alpha^{\act}$ is a $\Hom(G/A,\mathbb C^\times)$-valued 2-cocycle of $G$.
	We finally define a family of isomorphisms $\alpha^\act_{\mathbb C_g,\mathbb C_h,M} = \alpha^\act(g,h)(M) \cdot {\rm id}_{\msf p(gh \msf r(M))} : (\mathbb C_g \otimes \mathbb C_h) \act M \xrightarrow{\sim} \mathbb C_g \act (\mathbb C_h \act M)$, for any $g,h \in G$ and $M \in G/A$.
	Putting everything together, we obtain that the triple $(\mc M,\act,\alpha^{\act})$ defines a (left) $\Vect_G$-module category such that the pentagon axiom is ensured by the 2-cocycle condition ${\rm d}\alpha^\act=1$ . Applying similar techniques, we can construct $\Vect_G^\pi$-module categories from pairs $(A
	\subset G, \phi \in C^2(G, \rU(1)))$ such that $\alpha^{-1} {\sss |}_A = {\rm d}\phi$.
\end{example}

\noindent
Given a finite semi-simple left $\mc C$-module $\mc M$, we denote the set of representatives of isomorphism classes of simple objects in $\mc M$ via $\sOb(\mc M)$ such that $A \act M \simeq \bigoplus_{N \in {\rm sOb(\mc M)}} d^{AM}_N \, N$ with $d^{AM}_N := {\rm dim}_{\mathbb C}\Hom_{\mc M}(A \act M, N)$. Given any simple objects $A \in \sOb(\mc C)$ and $M,N \in \sOb(\mc M)$ such that $d^{AM}_N > 0$, we denote by $\{ |{}^{AM}_N,i \ra \, | \, i=1, \ldots, d^{AM}_N\}$ a choice of basis for $\Hom_{\mc M}(A \act M, N)$. Considering the compositions of morphisms
\begin{gather*}
	(A \otimes B) \act M 
	\xrightarrow{|{}^{AB}_C,i \ra \act {\rm id}_{M}}
	C \act M
	\xrightarrow{|{}^{CM}_N,j \ra} N
	\\
	(A \otimes B) \act M
	\xrightarrow{\alpha^{\act}_{A,B,M}}
	A \act (B \act M)
	\xrightarrow{{\rm id}_A \act |{}^{BM}_O,m \ra}
	A \act O
	\xrightarrow{|{}^{AO}_N,n \ra} N \, ,
\end{gather*}
for any simple objects $A,B,C \in \sOb(\mc C)$ and $M,N,O \in \sOb(\mc M)$, we realise that the module associator $\alpha^{\act}$ boils down to a collection of complex matrices
\begin{equation}
    \label{eq:SixJ}
    \SixJAct{A}{B}{C}{M}{N}{O}:
    \Hom_{\mc C}(A \otimes B,C) \otimes \Hom_{\mc M}(C \act M,N) \xrightarrow{\sim} \Hom_{\mc M}(B \act M, O) \otimes \Hom_{\mc M}(A \act O,N) \, ,
\end{equation}
whose entries are constants $\SixJAct{A}{B}{C}{M}{N}{O}_{ij,mn} \in \mathbb C$. Taking $\mc M= \mc C$ yields analogous matrices denoted by $\SixJ{A}{B}{C}{D}{E}{F}$ for the monoidal associator $\alpha$ in $\mc C$. Finally, it follows from the pentagon axiom that these constants satisfy the following equations
\begin{equation*}
    \label{eq:SixJEq}
    \sum_{m = 1}^{d^{DO}_N}
    \SixJAct{D}{C}{E}{M}{N}{O}_{jk,lm} 
    \SixJAct{A}{B}{D}{O}{N}{P}_{im,no} 
    =
    \sum_{F \in \sOb(\mc C)} \sum_{p=1}^{d^{BC}_F} \sum_{q=1}^{d^{AF}_E} \sum_{r=1}^{d^{FM}_P}
    \SixJ{A}{B}{D}{C}{E}{F}_{ij,pq}
    \SixJAct{A}{F}{E}{M}{N}{P}_{qk,ro}
    \SixJAct{B}{C}{F}{M}{P}{O}_{pr,ln} \, ,
\end{equation*}
for any choice of simple objects $A,B,C,D,E,F \in \sOb(\mc C)$ and $M,N,O,P \in \sOb(\mc M)$, and basis states in the relevant hom-sets.

Analogous to a left module category, we can define a notion of \emph{right} module category over a monoidal category in terms of a binary functor $\cat: \mc M \times \mc C \to \mc M$ and a natural isomorphism $\alpha^\cat: (\bul \cat \bul)\cat \bul \xrightarrow{\sim} \bul \cat (\bul \otimes \bul)$. In particular, a left module category over a monoidal category $\mc C$ defines a right module category over $\mc C^{\rm op}$. Combining the notions of left- and right-module categories yields the concept of bimodule category:
\begin{definition}[Bimodule category]
    Given a pair of monoidal categories $(\mc C, \mc D)$, a $(\mc C, \mc D)$-bimodule category is defined as a sextuple $(\mc M, \act, \cat, \alpha^\act, \alpha^\cat, \alpha^{\caact})$ such that $(\mc M, \act, \alpha^\act)$ defines a left $\mc C$-module category, $(\mc M, \cat, \alpha^\cat)$ defines a right $\mc C$-module category, and $\alpha^{\caact}: (\bul \act \bul) \cat \bul \xrightarrow{\sim} \bul \act (\bul \cat \bul)$ is a natural isomorphism satisfying two `pentagon axioms' involving either $\alpha^\act$ or $\alpha^\cat$. The natural isomorphism $\alpha^{\caact}$ shall be referred to as the bimodule associator and its components read $\alpha^{\caact}_{A,M,X}: (A \act M) \cat X \xrightarrow{\sim} A \act (M \cat X)$ for all $A \in \Ob(\mc C)$, $M \in \Ob(\mc M)$ and $X \in \Ob(\mc D)$.
\end{definition}

\noindent
Following the same steps as previously, we find that given a $(\mc C, \mc D)$-bimodule $\mc M$, the bimodule associator $\alpha^{\caact}$ boils down to a collection of complex matrices
\begin{equation}
    \label{eq:SixJBi}
    \SixJCaact{A}{M}{N}{X}{O}{P}:
    \Hom_{\mc M}(A \act M,N) \otimes \Hom_{\mc M}(N \cat X,O) \xrightarrow{\sim} \Hom_{\mc M}(M \cat X, P) \otimes \Hom_{\mc M}(A \act P,O) \, 
\end{equation}
whose entries are constants $\SixJCaact{A}{M}{N}{X}{O}{P}_{ij,mn} \in \mathbb C$. It then follows from the pentagon axiom involving $\alpha^{\act}$ and $\alpha^{\caact}$ that these constants satisfy the following equations
\begin{equation}
    \label{eq:SixJBiEq}
    \sum_{m = 1}^{d^{CP}_O}
    \SixJCaact{C}{M}{N}{X}{O}{P}_{jk,lm} 
    \SixJAct{A}{B}{C}{P}{O}{Q}_{im,no} 
    =
    \sum_{R \in \sOb(\mc M)} \sum_{p=1}^{d^{BM}_R} \sum_{q=1}^{d^{AR}_N} \sum_{r=1}^{d^{RX}_Q}
    \SixJAct{A}{B}{C}{M}{N}{R}_{ij,pq}
    \SixJCaact{A}{R}{N}{X}{O}{Q}_{qk,ro}
    \SixJCaact{B}{M}{R}{X}{Q}{P}_{pr,ln} \, ,
\end{equation}
for any choice of simple objects $A,B,C \in \sOb(\mc C)$, $M,N,O,P,Q,R \in \sOb(\mc M)$ and $X \in \sOb(\mc D)$, and basis states in the relevant hom-sets. The pentagon axiom involving $\alpha^{\cat}$ and $\alpha^{\caact}$ yields similar equations in terms of complex matrices 
\begin{equation}
	\label{eq:SixJCat}
	\SixJCat{M}{A}{O}{B}{N}{C}:     \Hom_{\mc M}(M \cat A ,O) \otimes \Hom_{\mc M}(O \cat B,N) \xrightarrow{\sim} \Hom_{\mc C}(A \otimes B,C) \otimes \Hom_{\mc M}(M \cat C,N) \, .
\end{equation}
We further require the notions of functor between module categories and morphisms between such functors:
\begin{definition}[Module category functor] 
    Given a monoidal category $\mc C$ and a pair of left $\mc C$-module categories $(\mc M, \act, \alpha^{\act})$ and $(\mc N,\actb,\alpha^{\actb})$, a $\mc C$-module functor is defined as a pair $(F,\omega)$ that consists of a functor $F:\mc M \to \mc N$ and a natural isomorphism $\omega: F(\bul \act \bul) \xrightarrow{\sim} \bul \actb F(\bul)$, whose components read $\omega_{A,M}: F(A \act M)\xrightarrow{\sim} A \actb F(M)$ for all $A \in \Ob(\mc C)$ and $M \in \Ob(\mc M)$, satisfying a `pentagon axiom' involving both $\alpha^{\act}$ and $\alpha^{\actb}$.
\end{definition}

\begin{definition}[Module category natural transformation]
    Given a monoidal category $\mc C$ and a pair of left $\mc C$-module functors $(F,\omega)$ and $(F',\omega')$, a $\mc C$-module transformation between $(F,\omega)$ and $(F',\omega')$ is a natural transformation $\theta: F \Rightarrow F'$ satisfying $\omega_{A,M} \circ ({\rm id}_A \actb \theta_M) = \theta_{A \act M} \circ \omega'_{A,M}$ for all $A \in \Ob(\mc C)$ and $M \in \Ob(\mc M)$. 
\end{definition}
\noindent
These notions can be readily adapted to the case of right module categories.
Given a monoidal category $\mc C$ and a pair $(\mc M, 
\mc N)$ of $\mc C$-module categories, we shall refer to $\Fun_{\mc C}(\mc M, \mc N)$ as the category whose objects are $\mc C$-module functors and 1-morphisms are $\mc C$-module natural transformations. In particular, it is useful to note that $\Fun_{\mc C}(\mc C, \mc C) \cong \mc C^{\rm op}$. Physically, given a (2+1)d topological lattice model with input (spherical fusion) category $\mc C$, the category $\Fun_{\mc C}(\mc M, \mc N)$ encodes the boundary operators at the interface of two gapped boundaries labelled by $\mc C$-module categories $\mc M$ and $\mc N$, respectively.

\bigskip \noindent
Let us now go up one level of abstraction. Given a ($\mathbb C$-linear strict) \emph{2-category} $\mc C$, we notate the set of objects via ${\rm Ob}(\mc C)$ and the category of morphisms (hom-category) between two objects $A,B \in \Ob(\mc C)$ via $\HomC_{\mc C}(A,B)$.\footnote{Notice that we use a different font in order to facilitate the distinction between hom-categories ($\HomC$) and hom-sets ($\Hom$).} As customary, we shall refer to objects and morphisms in hom-categories as 1- and 2-morphisms, respectively. For each triple $A,B,C \in \Ob(\mc C)$ the horizontal composition functor of 1- and 2-morphisms is written as $\circ : \HomC_{\mc C}(A,B) \times \HomC_\mc C(B,C) \to \HomC_\mc C(A,C)$ and the corresponding identity 1- and 2-morphisms are notated via ${\rm id}_A$ and ${\rm id}_{\rm id_A}$, respectively. The vertical composition rule of 2-morphisms in hom-categories is written as $\cdot$ and the corresponding identity 1-morphisms as ${\rm id}_f$ for any $f \in \Ob(\HomC_\mc C(A,B))$. Henceforth, we assume that every $2$-category is $\mathbb C$-linear.

We shall think of a monoidal 2-category  as a monoidal bicategory in the sense of  \cite{gurski2011loop}, whose underlying bicategory is a strict 2-category and such that the unital conditions are strictly satisfied.\footnote{In particular, the \emph{interchanger} 2-isomorphisms are taken to be identities.} Concretely, we think of a monoidal  2-category as a quintuple $\mc C \equiv (\mc C , \otimes , \mathbb 1, \alpha, \pi)$ consisting of a 2-category $\mc C$, a monoidal binary 2-functor $\mc C \times \mc C \to \mc C$, a distinguished unit object $\mathbb 1 \in \Ob(\mc C)$, and an adjoint 2-natural equivalence $\alpha: (\bul \otimes \bul) \otimes \bul \to \bul \otimes (\bul \otimes \bul)$  satisfying the `pentagon axiom' up to an invertible modification $\pi$ fulfilling the `associahedron axiom' in $\Hom_{\HomC(\mc C)}((((\bul  \bul)  \bul )  \bul) \bul, \bul (\bul  (\bul (\bul \bul))))$. The invertible modification $\pi$ is referred to as the monoidal \emph{pentagonator} and its components read $\pi_{A,B,C,D}: (\alpha_{A,B,C}\otimes {\rm id}_D) \circ \alpha_{A,B \otimes C,D} \circ ({\rm id}_A \otimes \alpha_{B,C,D}) \Rightarrow \alpha_{A \otimes B , C, D} \circ \alpha_{A,B,C\otimes D}$.

Referring to finite, $\mathbb C$-linear, abelian, semi-simple $\Vect$-module categories as \emph{2-vector spaces}, a prototypical example of a monoidal 2-category is the category $\TVect$ of 2-vector spaces, $\Vect$-module functors and $\Vect$-module natural transformations. This 2-category has a single simple object, namely the (fusion) category $\Vect$ thought as a module category over itself, and the monoidal functor is provided by the Deligne tensor product of abelian categories. Thinking of $\TVect$ as a categorification of $\Vect$, a more sophisticated example can be obtained as a categorification of the notion of group-graded vector spaces by promoting $\Vect$ to $\TVect$:
\begin{example}[Group-graded 2-vector spaces\label{ex:2vec}]
    Given a finite group $G$ and a normalised group 4-cocycle $\pi$ in $H^4(G, {\rm U}(1))$, we define the monoidal 2-category $\VG$ as the category whose objects are $G$-graded 2-vector spaces of the form $\msf V = \bigboxplus_{g \in G} \msf V_g$, 1-morphisms are grading preserving $\Vect$-module functors, and 2-morphisms are $\Vect$-module natural transformations.  The monoidal 2-functor is defined on homogeneous components in terms of the Deligne tensor product $\boxtimes$ of abelian categories as
    \begin{equation*}
        (\msf V \boxtimes \msf W)_g = \bigboxplus_{x \in G}\msf V_x \boxtimes \msf V_{x^{-1}g} 
    \end{equation*}
    with unit $\mathbb 1$ such that $\mathbb 1_g = \delta_{g,\mathbb1_G}\Vect$. The monoidal associator evaluates to the identity 1-morphism, whereas the  pentagonator is provided via $\pi_{\msf T_g,\msf U_h,\msf V_k,\msf W_l} = \pi(g,h,k,l) \cdot (\mathring{\pi}_{\msf T,\msf U,\msf V,\msf W})_{ghkl}$, 
    for all $g,h,k,l \in G$, where $\mathring{\pi}$ refers to the pentagonator in $\TVect$. The $|G|$-many simple objects are the categories $\Vect_g$, for every $g \in G$, satisfying $\HomC_{\VG}(\Vect_g,\Vect_h) \cong \delta_{g,h} \Vect$ and $\Vect_g \boxtimes \Vect_h \cong \Vect_{gh}$.
\end{example}

\noindent
Generally, examples of monoidal 2-categories can be conveniently constructed as the 2-categories $\MOD(\mc C)$ of module categories, module category functors and module category natural transformations over monoidal categories $\mc C$. In particular, the monoidal 2-category $\TRep(G) := \MOD(\Vect_G^{\rm op})$ of `2-representations of $G$' will play a prominent role in the following. 

Categorifying the notion of module over a monoidal category yields the concept of \emph{module 2-category}: 
\begin{definition}[Left module 2-category\label{def:mod2Cat}] 
    Given a monoidal 2-category $\mc C \equiv (\mc C, \otimes, \mathbb 1, \alpha,\pi)$, we define a left module 2-category over $\mc C$ as a quadruple $(\mc M, \act,\alpha^{\triangleright}, \pi^{\act})$ that consists of a 2-category $\mc M$, a binary action 2-functor $\act: \mc C \times \mc M \to \mc M$, an adjoint 2-natural equivalence $\alpha^\act : (\bul \otimes \bul) \act \bul \to \bul \act (\bul \act \bul)$ satisfying the `pentagon axiom' up to an invertible modification $\pi^\act$ whose components are defined via
    \begin{align*}
    	\begin{tikzcd}[ampersand replacement=\&, column sep=4em, row sep=2.5em]
    		|[alias=A1]|((A \otimes B) \otimes C ) \act M
    		\&
    		|[alias=B1]|(A \otimes B) \act (C \act M)
    		\&
    		|[alias=C1]| A \act (B \act ( C \act M))
    		\\
    		|[alias=A2]|(A \otimes (B \otimes C)) \act M
    		\&
    		\&
    		|[alias=C2]| A \act ((B \otimes C) \act M)
    		\arrow[from=A1,to=B1,"\alpha^\act_{A \otimes B,C,M}"]
    		\arrow[from=B1,to=C1,"\alpha^\act_{A,B,C \act M}"]
    		\arrow[from=C2,to=C1,"{\rm id}_A \act \alpha^\act_{B,C,M}"']
    		\arrow[from=A1,to=A2,"\alpha_{A,B,C}\act {\rm id}_M"]
    		\arrow[from=A2,to=C2,"\alpha^\act_{A,B \otimes C,M}"',""{name=X,above}]
            \arrow[Rightarrow,from=X,to=B1,"\pi^\act_{A,B,C,M}"',shorten <= 0.5em, shorten >= 0.5em]
    	\end{tikzcd}
    \end{align*}
    for all $A,B,C \in \Ob(\mc C)$ and $M \in \Ob(\mc M)$. The invertible modification $\pi^\act$, which shall be referred to as the left module pentagonator, is required to fulfil an `associahedron' coherence axiom encoded into the equality of the commutative diagrams
    \begin{align*}
    	\begin{tikzcd}[ampersand replacement=\&, column sep=1.5em, row sep=2.5em ]
    	    |[alias=A1]| (\bul\bul)\act(\bul \act (\bul \act \bul)) \&\&
    	    |[alias=C1]| (\bul\bul) \act ((\bul\bul)\act \bul) \&\&
    	    |[alias=E1]| \bul \act (\bul \act ((\bul \bul)\act \bul))
    	    \\
            |[alias=A2]| ((\bul \bul)\bul)\act (\bul \act \bul) \& 
            |[alias=B2]| (((\bul \bul)\bul)\bul)\act \bul \& 
            |[alias=C2]| ((\bul \bul)(\bul \bul))\act \bul \& 
            |[alias=D2]| (\bul(\bul(\bul \bul)))\act \bul \& 
            |[alias=E2]| \bul \act ((\bul(\bul \bul))\act \bul)
            \\
            |[alias=A3]| (\bul(\bul\bul))\act(\bul \act \bul) \&
            |[alias=B3]| ((\bul(\bul\bul))\bul) \act \bul \&\&
            |[alias=D3]| (\bul((\bul\bul)\bul))\act \bul \&
            |[alias=E3]| \bul \act (((\bul\bul)\bul)\act \bul)
            \arrow[from=B2,to=A2,"\alpha^\act"',pos=0.35]
            \arrow[from=B2,to=C2,"\alpha \act {\rm id}"]
            \arrow[from=C2,to=D2,"\alpha \act {\rm id}"]
            \arrow[from=D2,to=E2,"\alpha^\act",""{name=Y,above}]
            \arrow[from=B3,to=A3,"\alpha^\act",pos=0.35]
            \arrow[from=B3,to=D3,"\alpha \act {\rm id}"',""{name=X,above}]
            \arrow[from=D3,to=E3,"\alpha^\act"']
            \arrow[from=C1,to=A1,"{\rm id} \act \alpha^\act"']
            \arrow[from=C1,to=E1,"\alpha^\act"]
            \arrow[from=A2,to=A3,"\alpha \act {\rm id}",""{name=M,right}]
            \arrow[from=B2,to=B3,"\alpha \act {\rm id}",""{name=N,left}]
            \arrow[from=D3,to=D2,"\alpha \act {\rm id}"',""{name=O,right}]
            \arrow[from=E3,to=E2,"{\rm id} \act \alpha"',""{name=P,left}]
            \arrow[from=A2,to=A1,"\alpha^\act"']
            \arrow[from=C2,to=C1,"\alpha^\act"',""{name=Z,left}]
            \arrow[from=E2,to=E1,"{\rm id} \act \alpha^\act"']
            \arrow[Rightarrow,from=Z,to=A2,"\pi^\act",sloped, bend right=15,shorten <= 0.5em, shorten >= 0.5em]
            \arrow[Rightarrow,from=X,to=C2,"\pi \act {\rm id}"',shorten <= 0.5em, shorten >= 0.5em]
            \arrow[Rightarrow,from=Y,to=C1,"\pi^\act",sloped,bend right=10,shorten <= 1em, shorten >= 0.5em]
            \arrow[from=M,to=N,Rightarrow,white,"{\color{black}\cong}" description, shorten <= 4em, shorten >= 2em]
            \arrow[from=O,to=P,Rightarrow,white,"{\color{black}\cong}" description, shorten <= 2em, shorten >= 2em]
    	\end{tikzcd} \phantom{,}
    \end{align*}
    and
    \begin{align*}
    	\begin{tikzcd}[ampersand replacement=\&, column sep=1.5em, row sep=2.5em ]
    	    |[alias=A1]| (\bul\bul)\act(\bul \act (\bul \act \bul)) \&\&
    	    |[alias=C1]| (\bul\bul) \act ((\bul\bul)\act \bul) \&\&
    	    |[alias=E1]| \bul \act (\bul \act ((\bul \bul)\act \bul))
    	    \\[-1.38em]
    	    \&\& |[alias=C15]| \bul \act (\bul \act (\bul \act (\bul \act \bul))) \&\&
    	    \\[-2.8em]
            |[alias=A2]| ((\bul \bul)\bul)\act (\bul \act \bul) \&\&\&\&
            |[alias=E2]| \bul \act ((\bul(\bul \bul))\act \bul)
            \\[-2.8em]
            \&\& |[alias=C25]| \bul \act ((\bul \bul)\act (\bul\act \bul)) \&\&
            \\[-1.38em]
            |[alias=A3]| (\bul(\bul\bul))\act(\bul \act \bul) \&
            |[alias=B3]| ((\bul(\bul\bul))\bul) \act \bul \&\&
            |[alias=D3]| (\bul((\bul\bul)\bul))\act \bul \&
            |[alias=E3]| \bul \act (((\bul\bul)\bul)\act \bul)
            \arrow[from=B3,to=A3,"\alpha^\act",pos=0.35]
            \arrow[from=B3,to=D3,"\alpha \act {\rm id}"']
            \arrow[from=D3,to=E3,"\alpha^\act"',""{name=Z,above}]
            \arrow[from=C1,to=A1,"{\rm id} \act \alpha^\act"']
            \arrow[from=C1,to=E1,"\alpha^\act"]
            \arrow[from=A2,to=A3,"\alpha \act {\rm id}"]
            \arrow[from=E3,to=E2,"{\rm id} \act \alpha"']
            \arrow[from=A2,to=A1,"\alpha^\act"']
            \arrow[from=E2,to=E1,"{\rm id} \act \alpha^\act"',""{name=Y,left}]
            \arrow[from=A3,to=C25,"\alpha^\act",sloped,""{name=X,above}]
            \arrow[from=C25,to=C15,"{\rm id} \act \alpha^\act"]
            \arrow[from=E3,to=C25,"{\rm id} \act \alpha^\act",sloped]
            \arrow[from=A1,to=C15,"\alpha^\act",sloped]
            \arrow[from=E1,to=C15,"{\rm id} \act \alpha^\act",sloped]
            \arrow[Rightarrow,from=X,to=A1,"\pi^\act"',bend right=15, shorten <= 1em, shorten >= 0.5em]
            \arrow[Rightarrow,from=Y,to=C25,"{\rm id}\act \pi^\act"',sloped,bend right=10,shorten <= 0.5em, shorten >= 0.5em]
            \arrow[yshift=-2pt,Rightarrow,from=E3,to=A3,"\pi^\act", bend right=7, shorten <= 5em, shorten >= 5em]
            \arrow[from=C1,to=C15,Rightarrow,white,"{\color{black}\cong}" description]
    	\end{tikzcd} \!\!\!\! ,
    \end{align*}
    where we omitted the monoidal 2-functor $\otimes$ and used the placeholder $\bul$ for convenience.
\end{definition}

\noindent
Henceforth, we implicitly assume that every monoidal or module 2-category is finite and semi-simple in the sense of Douglas and Reutter \cite{douglas2018fusion}.  
Akin to the lower-categorical setting, every monoidal 2-category $\mc C$ defines a (left) module 2-category over itself, referred to as the \emph{regular} $\mc C$-module 2-category, such that the action 2-functor is provided by the monoidal structure in $\mc C$. A more interesting family of examples goes as follows:

\begin{example}[$\TVect_G$-module 2-categories\label{ex:TVecGModule}] 
    Let $A \subset G$ be a subgroup of $G$ and $\lambda$ a normalised group 3-cocycle in $H^3(A, \rU(1))$. Following the notations of ex.~\ref{ex:VecGModule}, we consider the 2-category $\mc M$ whose collection of simple objects is the set $G/A$ of left cosets and define the binary functor $\act : \TVect_G \times \mc M \to \mc M$ via $\Vect_g \act M := \msf p(g \msf r(M))$, for all $g \in G$ and $M \in G /A$. We notate via $a_{g,M}$ the group element in $A$ satisfying $g \msf r(M) =  \msf r (\Vect_g \act M) a_{g,M}$, for all $g \in G$ and  $M \in G /A$. We consider the 3-cochain $\pi^{\act} \in C^3(G, \Hom(G/A,\mathbb C^\times))$ defined by $\pi^{\act}(g,h,k)(M) := \lambda(a_{g,\Vect_{hk}\act M}, a_{h,\Vect_k\act M},a_{k,M})$ for any $g,h,k \in G$ and $\,M \in G/A$. It follows from the equation satisfied by $a_{\bul,\bul}$ as well as the cocycle condition ${\rm d}\lambda = 1$ that
    \begin{align}
        \pi^{\act}(h,k,l)(M)\, \pi^{\act}(g,hk,l)(M) \, \pi^{\act}(g,h,k)(\Vect_l \act M) = \pi^{\act}(gh,k,l)(M) \, \pi^{\act}(g,h,kl)(M) \, ,
    \end{align}
    for any $g,h,k,l \in G$ and $M \in G/A$,
    i.e. $\pi^{\act}$ is a $\Hom(G/A,\mathbb C^\times)$-valued 3-cocycle of $G$.
    We finally define a family of 2-isomorphisms $\pi^\act_{\Vect_g,\Vect_h,\Vect_k,M}$ given by $\pi^{\act}(g,h,k)(M)$  times the canonical pentagonator in $\TVect$, for any $g,h,k \in G$ and $M \in G/A$.
    Putting everything together, we obtain that the triple $(\mc M,\act,{\rm id},\pi^{\act})$ defines a (left) $\TVect_G$-module 2-category such that the pentagonator axiom is ensured by the 3-cocycle condition ${\rm d}\pi^\act=1$ . Applying similar techniques, we can construct $\VG$-module categories from pairs $(A 
    \subset G, \lambda \in C^3(G, \rU(1)))$ such that $\pi^{-1} {\sss |}_A = {\rm d}\lambda$. 
\end{example}

\noindent 
We can similarly define the notion of right module 2-category in terms of a right module associator fulfilling the `pentagon axiom' up to an invertible modification $\pi^\cat$, which combined with that of left module 2-category yields the concept of bimodule 2-category:
\begin{definition}[Bimodule 2-category\label{def:biMod2Cat}]
    Given a pair of monoidal 2-categories $(\mc C, \mc D)$, we define a $(\mc C, \mc D)$-bimodule 2-category as a decuple $(\mc M, \act, \cat, \alpha^\act, \alpha^\cat, \alpha^{\caact},\pi^\act,\pi^\cat,\pi^{\caact}, \tilde \pi^{\caact})$ in such a way that $(\mc M, \act, \alpha^\act,\pi^\act)$ defines a left $\mc C$-module 2-category, $(\mc M, \cat, \alpha^\cat, \pi^\cat)$ defines a right $\mc C$-module 2-category, and $\alpha^{\caact}: (\bul \act \bul) \cat \bul \to \bul \act (\bul \cat \bul)$ is an adjoint 2-natural equivalence satisfying the defining `pentagon axioms' of a bimodule category involving $\alpha^\act$ or $\alpha^\cat$ up to invertible modifications $\pi^{\caact}$ and $\tilde \pi^{\caact}$, respectively, whose components are defined via
    \begin{gather*}
        \begin{tikzcd}[ampersand replacement=\&, column sep=4em, row sep=2.5em]
            |[alias=A1]|((A \otimes B) \act M ) \cat X
            \&
            |[alias=B1]|(A \otimes B) \act (M \cat X)
            \&
            |[alias=C1]| A \act (B \act ( M \cat X))
            \\
            |[alias=A2]|(A \act (B \act M)) \cat X
            \&
            \&
            |[alias=C2]| A \act ((B \act M) \cat X)
            \arrow[from=A1,to=B1,"\alpha^{\caact}_{A \otimes B,M,X}"]
            \arrow[from=B1,to=C1,"\alpha^\act_{A,B,M \cat X}"]
            \arrow[from=C2,to=C1,"{\rm id}_A \act \alpha^{\caact}_{B,M,X}"']
            \arrow[from=A1,to=A2,"\alpha^\act_{A,B,M}\cat {\rm id}_X"]
            \arrow[from=A2,to=C2,"\alpha^{\caact}_{A,B \act M,X}"',""{name=X,above,pos=0.505}]
            \arrow[from=X,to=B1,Rightarrow,"\pi^{\caact}_{A,B,M,X}"',shorten <= 0.5em, shorten >= 0.5em]
        \end{tikzcd}
        	\\
        \begin{tikzcd}[ampersand replacement=\&, column sep=4em, row sep=2.5em]
            |[alias=A1]|((A \act M) \cat X ) \cat Y
            \&
            |[alias=B1]|(A \act M) \cat (X \otimes Y)
            \&
            |[alias=C1]| A \act (M \cat (X \otimes Y))
            \\
            |[alias=A2]|(A \act (M \cat X))\cat Y
            \&
            \&
            |[alias=C2]| A \act ((M \cat X)\cat Y)
            \arrow[from=A1,to=B1,"\alpha^\cat_{A \act M,X,Y}"]
            \arrow[from=B1,to=C1,"\alpha^{\caact}_{A,M,X \otimes Y }"]
            \arrow[from=C2,to=C1,"{\rm id}_A \act \alpha^\cat_{M,X,Y}"']
            \arrow[from=A1,to=A2,"\alpha^{\caact}_{A,M,X}\cat {\rm id}_Y"]
            \arrow[from=A2,to=C2,"\alpha^{\caact}_{A,M \cat X,Y}"',""{name=X,above,pos=0.497}]
            \arrow[Rightarrow,from=X,to=B1,"\tilde \pi^{\caact}_{A,M,X,Y}"',shorten <= 0.5em, shorten >= 0.5em]
        \end{tikzcd}
    \end{gather*}
    for all $A,B \in \Ob(\mc C)$, $M \in \Ob(\mc M)$ and $X,Y \in \Ob(\mc D)$. The invertible modifications $\pi^{\caact}$ and $\tilde \pi^{\caact}$ shall be referred to as the bimodule left and right pentagonators, respectively. They are each required to fulfil an `associahedron axiom', akin to those of the module pentagonators, involving either $\pi^\act$ or $\pi^{\cat}$ in $\Hom_{\HomC(\mc M)}( (((\bul \bul)\bul)\act \bul)\cat \bul, \bul \act (\bul \act (\bul \act (\bul \cat \bul))) )$ and $\Hom_{\HomC(\mc M)}( (((\bul \act \bul) \cat \bul)\cat \bul)\cat \bul , \bul \act (\bul \cat (\bul(\bul \bul))) )$, respectively, and they satisfy together a third `associahedron axiom' in $\Hom_{\HomC(\mc M)}((((\bul \bul) \act \bul)\cat \bul ) \cat \bul, \bul \act ( \bul \act (\bul \cat (\bul \bul))))$.
\end{definition}

\noindent
Proceeding with our categorification, we consider the notions of 2-functor between module 2-categories and morphisms between such 2-functors:
\begin{definition}[Module 2-category 2-functor\label{def:mod2Cat2Fun}]
    Given a monoidal 2-category $\mc C$ and a pair of left $\mc C$-module 2-categories $(\mc M, \act, \alpha^{\act},\pi^{\act})$ and $(\mc N,\actb,\alpha^{\actb},\pi^{\actb})$, we define a $\mc C$-module 2-functor as a triple $(F,\omega,\Omega)$ that consists of a (strict) 2-functor $F:\mc M \to \mc N$ and an adjoint 2-natural equivalence $\omega: F(\bul \act \bul) \to \bul \actb F(\bul)$ satisfying a `pentagon axiom' up to an invertible modification $\Omega$ whose components are defined via
    \begin{align*}
    	\begin{tikzcd}[ampersand replacement=\&, column sep=4em, row sep=2.5em]
    		|[alias=A1]|F((A \otimes B) \act M )
    		\&
    		|[alias=B1]|(A \otimes B) \actb F(M)
    		\&
    		|[alias=C1]| A \actb (B \actb F(M))
    		\\
    		|[alias=A2]| F(A \act (B \act M))
    		\&
    		\&
    		|[alias=C2]| A \actb F(B \act M)
    		\arrow[from=A1,to=B1,"\omega_{A \otimes B,M}"]
    		\arrow[from=B1,to=C1,"\alpha^{\actb}_{A,B,F(M)}"]
    		\arrow[from=C2,to=C1,"{\rm id}_A \actb \omega_{B,M}"']
    		\arrow[from=A1,to=A2,"F(\alpha^{\act}_{A,B,M})"]
    		\arrow[from=A2,to=C2,"\omega_{A,B \act M}"',""{name=X,above,pos=0.495}]
            \arrow[Rightarrow,from=X,to=B1,"\Omega_{A,B,M}"',shorten <= 0.5em, shorten >= 0.5em]
    	\end{tikzcd}
    \end{align*}
    for all $A,B \in \Ob(\mc C)$ and $M \in \Ob(\mc M)$. The invertible modification $\Omega$ is required to fulfil a coherence axiom encoded into the equality of the commutative diagrams
    \begin{align*}
    	\begin{tikzcd}[ampersand replacement=\&, column sep=1.5em, row sep=2.5em ]
    	    |[alias=A1]| (\bul\bul)\actb (\bul \actb F(\bul)) \&\&
    	    |[alias=C1]| (\bul\bul) \actb F(\bul \act \bul) \&\&
    	    |[alias=E1]| \bul \actb (\bul \actb F(\bul \act \bul))
    	    \\
            |[alias=A2]| ((\bul \bul)\bul)\actb F(\bul) \& 
            |[alias=B2]| F(((\bul \bul) \bul) \act \bul) \& 
            |[alias=C2]| F((\bul \bul)\act (\bul \act \bul)) \& 
            |[alias=D2]| F(\bul \act(\bul \act (\bul \act \bul))) \& 
            |[alias=E2]| \bul \actb F(\bul \act (\bul \act \bul))
            \\
            |[alias=A3]| (\bul(\bul\bul))\actb F(\bul)\&
            |[alias=B3]| F((\bul(\bul\bul))\act \bul) \&\&
            |[alias=D3]| F(\bul \act ((\bul\bul)\act \bul)) \&
            |[alias=E3]| \bul \actb F((\bul \bul)\act \bul)
            \arrow[from=B2,to=A2,"\omega"',pos=0.35]
            \arrow[from=B2,to=C2,"F(\alpha^\act)"]
            \arrow[from=C2,to=D2,"F(\alpha^\act)"]
            \arrow[from=D2,to=E2,"\omega",""{name=Y,above}]
            \arrow[from=B3,to=A3,"\omega",pos=0.35]
            \arrow[from=B3,to=D3,"F(\alpha^{\act})"',""{name=X,above,pos=0.48}]
            \arrow[from=D3,to=E3,"\omega"']
            \arrow[from=C1,to=A1,"{\rm id} \actb \omega"']
            \arrow[from=C1,to=E1,"\alpha^{\actb}"]
            \arrow[from=A2,to=A3,"\alpha \actb F(\rm id)",""{name=M,right}]
            \arrow[from=B2,to=B3,"F(\alpha \act {\rm id})",""{name=N,left}]
            \arrow[from=D3,to=D2,"F({\rm id} \act \alpha^\act)",""{name=O,right}]
            \arrow[from=E3,to=E2,"{\rm id}\actb F(\alpha^\act)",""{name=P,left}]
            \arrow[from=A2,to=A1,"\alpha^{\actb}"']
            \arrow[from=C2,to=C1,"\omega"',""{name=Z,left}]
            \arrow[from=E2,to=E1,"{\rm id} \actb \omega"']
            \arrow[Rightarrow,from=Z,to=A2,"\Omega",sloped, bend right=15,shorten <= 0.5em, shorten >= 0.5em]
            \arrow[Rightarrow,from=X,to=C2,"F(\pi^\act)"',shorten <= 0.5em, shorten >= 0.5em]
            \arrow[Rightarrow,from=Y,to=C1,"\Omega",sloped,bend right=10,shorten <= 1em, shorten >= 0.5em]
            \arrow[from=M,to=N,Rightarrow,white,"{\color{black}\cong}" description, shorten <= 4em, shorten >= 4em]
            \arrow[from=O,to=P,Rightarrow,white,"{\color{black}\cong}" description, shorten <= 4em, shorten >= 4em]
    	\end{tikzcd} \phantom{,}
    \end{align*}
    and
    \begin{align*}
    	\begin{tikzcd}[ampersand replacement=\&, column sep=1.5em, row sep=2.5em ]
    	    |[alias=A1]| (\bul\bul)\actb (\bul \actb F(\bul)) \&\&
    	    |[alias=C1]| (\bul\bul) \actb F(\bul \act \bul) \&\&
    	    |[alias=E1]| \bul \actb (\bul \actb F(\bul \act \bul))
    	    \\[-1.38em]
    	    \&\& |[alias=C15]| \bul \actb (\bul \actb (\bul \actb F(\bul))) \&\&
    	    \\[-2.8em]
            |[alias=A2]| ((\bul \bul)\bul)\actb F(\bul) \&\&\&\&
            |[alias=E2]| \bul \actb F(\bul \act (\bul \act \bul))
            \\[-2.8em]
            \&\& |[alias=C25]| \bul \actb ((\bul \bul)\actb F(\bul)) \&\&
            \\[-1.38em]
            |[alias=A3]| (\bul(\bul\bul))\actb F(\bul)\&
            |[alias=B3]| F((\bul(\bul\bul))\act \bul) \&\&
            |[alias=D3]| F(\bul \act ((\bul\bul)\act \bul)) \&
            |[alias=E3]| \bul \actb F((\bul \bul)\act \bul)
            \arrow[from=B3,to=A3,"\omega",pos=0.35]
            \arrow[from=B3,to=D3,"F(\alpha^\act)"']
            \arrow[from=D3,to=E3,"\omega"',""{name=Z,above}]
            \arrow[from=C1,to=A1,"{\rm id}\actb \omega"']
            \arrow[from=C1,to=E1,"\alpha^{\actb}"]
            \arrow[from=A2,to=A3,"\alpha \actb F(\rm id)"]
            \arrow[from=E3,to=E2,"{\rm id} \actb F(\alpha^\act)"]
            \arrow[from=A2,to=A1,"\alpha^{\actb}"']
            \arrow[from=E2,to=E1,"{\rm id} \actb \omega"',""{name=Y,left}]
            \arrow[from=A3,to=C25,"\alpha^{\actb}",sloped,""{name=X,above}]
            \arrow[from=C25,to=C15,"{\rm id}\actb \alpha^{\actb}"]
            \arrow[from=E3,to=C25,"{\rm id} \actb \omega",sloped]
            \arrow[from=A1,to=C15,"\alpha^{\actb}",sloped]
            \arrow[from=E1,to=C15,"{\rm id}\actb ({\rm id} \actb \omega) ",sloped]
            \arrow[Rightarrow,from=X,to=A1,"\pi^{\actb}"',bend right=15, shorten <= 1em, shorten >= 0.5em]
            \arrow[Rightarrow,from=Y,to=C25,"{\rm id}\actb \Omega"',sloped,bend right=10,shorten <= 0.5em, shorten >= 0.5em]
            \arrow[yshift=-2pt,Rightarrow,from=E3,to=A3,"\Omega", bend right=7, shorten <= 5em, shorten >= 5em]
            \arrow[from=C1,to=C15,Rightarrow,white,"{\color{black}\cong}" description]
    	\end{tikzcd} \! ,
    \end{align*}
    where we omitted the monoidal 2-functor $\otimes$ for convenience.
\end{definition}

\begin{definition}[Module 2-category 2-natural transformation\label{def:mod2Cat2Nat}]
    Given a monoidal 2-category $\mc C$ and a pair of (left) $\mc C$-module 2-functors $(F,\omega,\Omega)$ and $(F',\omega',\Omega')$, we define a $\mc C$-module 2-natural transformation from $(F,\omega,\Omega)$ to $(F',\omega',\Omega')$ as a tuple $(\theta,\Theta)$ such that $\theta: F \Rightarrow F'$ is a 2-natural transformation and $\Theta$ is an invertible modification whose components are defined via
    \begin{align*}
    	\begin{tikzcd}[ampersand replacement=\&, column sep=4em, row sep=2.5em]
    		|[alias=A1]|F(A \act M)
    		\&
    		|[alias=B1]|A \actb F(M)
    		\\
    		|[alias=A2]|F'(A \act M)
    		\&
    		|[alias=B2]| A \actb F'(M)
    		\arrow[from=A1,to=B1,"\omega_{A,M}"]
    		\arrow[from=B1,to=B2,"{\rm id}_A \actb \theta_M"]
    		\arrow[from=A1,to=A2,"\theta_{A \act M}"]
    		\arrow[from=A2,to=B2,"\omega'_{A,M}"']
            \arrow[Rightarrow,from=B1,to=A2,"\Theta_{A,M}",sloped,shorten <= 0.5em, shorten >= 0.5em]
    	\end{tikzcd}
    \end{align*}
    for all $A \in \Ob(\mc C)$ and $M \in \Ob(\mc M)$. The invertible modification $\Theta$ is required to fulfil a coherence axiom encoded into the equality of the commutative diagrams
    \begin{align*}
    	\begin{tikzcd}[ampersand replacement=\&, column sep=1.5em, row sep=2.5em ]
    	    \&
            |[alias=B1]| F'((\bul \bul)\act \bul) \&
            |[alias=C1]| (\bul \bul) \actb F'(\bul) \&
            |[alias=D1]| \bul \actb (\bul \actb F'(\bul)) \&
            \\
            |[alias=A2]| F'(\bul \act (\bul \act \bul)) \&
            |[alias=B2]| F((\bul \bul)\act \bul) \&
            |[alias=C2]| (\bul \bul) \actb F(\bul) \&
            |[alias=D2]| \bul \actb (\bul \actb F(\bul))
            \\
            \&
            |[alias=B3]| F(\bul \act (\bul \act \bul)) \&\&
            |[alias=D3]| \bul \actb F(\bul \act \bul)
            \arrow[from=B1,to=C1,"\omega'"]
            \arrow[from=C1,to=D1,"\alpha^{\actb}"]
            \arrow[from=B2,to=C2,"\omega"]
            \arrow[from=C2,to=D2,"\alpha^{\actb}"]
            \arrow[from=B2,to=B3,"F(\alpha^\act)"]
            \arrow[from=B3,to=D3,"\omega"',""{name=X,above,pos=0.48}]
            \arrow[from=D3,to=D2,"{\rm id}\actb \omega"']
            \arrow[from=B2,to=B1,"\theta"']
            \arrow[from=C2,to=C1,"{\rm id} \actb \theta"']
            \arrow[from=D2,to=D1,"{\rm id}\actb ({\rm id} \actb \theta)"']
            \arrow[from=B3,to=A2,"\theta"]
            \arrow[from=B1,to=A2,"F'(\alpha^\act)",sloped]
            \arrow[Rightarrow,from=C2,to=B1,"\Theta",sloped,shorten <= 0.5em, shorten >= 0.5em]
            \arrow[Rightarrow,from=X,to=C2,"\Omega"',shorten <= 0.5em, shorten >= 0.5em]
            \arrow[from=D2,to=C1,Rightarrow,white,"{\color{black}\cong}" description, shorten <= 4em, shorten >= 4em]
            \arrow[Rightarrow,from=B3,to=B1,"\theta_{\alpha^{\act}}\,",bend left=50, shorten <=1.5em, shorten >=1.5em]
    	\end{tikzcd} \phantom{\hspace{-1.7em} .}
    \end{align*}
    and
    \begin{align*}
    	\begin{tikzcd}[ampersand replacement=\&, column sep=1.5em, row sep=2.5em ]
    	    \&
            |[alias=B1]| F'((\bul \bul)\act \bul) \&
            |[alias=C1]| (\bul \bul) \actb F'(\bul) \&
            |[alias=D1]| \bul \actb (\bul \actb F'(\bul)) \&
            \\
            |[alias=A2]| F'(\bul \act (\bul \act \bul)) \&
            \& |[alias=C2]| \bul \actb F'(\bul \act \bul)\&
            |[alias=D2]| \bul \actb (\bul \actb F(\bul))
            \\
            \&
            |[alias=B3]| F(\bul \act (\bul \act \bul)) \&\&
            |[alias=D3]| \bul \actb F(\bul \act \bul)
            \arrow[from=B1,to=C1,"\omega'"]
            \arrow[from=C1,to=D1,"\alpha^{\actb}"]
            \arrow[from=B3,to=D3,"\omega"']
            \arrow[from=D3,to=D2,"{\rm id}\actb \omega"']
            \arrow[from=D2,to=D1,"{\rm id}\actb ({\rm id} \actb \theta)"']
            \arrow[from=B3,to=A2,"\theta"]
            \arrow[from=B1,to=A2,"F'(\alpha^\act)",sloped]
            \arrow[from=D3,to=C2,"{\rm id} \actb \theta",sloped]
            \arrow[from=C2,to=D1,"{\rm id} \actb \omega'",sloped]
            \arrow[from=A2,to=C2,"\omega'",""{name=X,above,pos=0.5}]
            \arrow[Rightarrow,from=X,to=C1,"\Omega'",sloped,shorten <= 0.8em, shorten >= 0.5em]
            \arrow[Rightarrow,from=D3,to=A2,"\Theta",sloped,shorten <= 0.5em, shorten >= 0.5em]
            \arrow[Rightarrow,from=D2,to=C2,"{\rm id}\actb \Theta"',shorten <= 0.1em, shorten >= 0.1em]
    	\end{tikzcd} \hspace{-1.7em} .
    \end{align*}
\end{definition}

\noindent
The final step of this categorification consists in introducing a notion of 2-morphism between morphisms of module 2-category 2-functors:

\begin{definition}[Module 2-category modification\label{def:mod2Cat2Mod}] 
    Given a monoidal 2-category $\mc C$ and a pair of $\mc C$-module 2-natural transformations $(\theta,\Theta)$ and $(\theta',\Theta')$, we define a $\mc C$-module modification from $(\theta,\Theta)$ to $(\theta',\Theta')$ as a modification $\vartheta: \theta \Rrightarrow \theta'$ such that the diagram
    \begin{align*}
    	\begin{tikzcd}[ampersand replacement=\&, column sep=6em, row sep=2.5em]
    		|[alias=A1]| \omega_{A,M}\circ ({\rm id}_A \actb \theta_M)
    		\&
    		|[alias=B1]| \omega_{A,M} \circ ({\rm id}_A \actb \theta'_M)
    		\\
    		|[alias=A2]| \theta_{A \act M} \circ \omega'_{A,M}
    		\&
    		|[alias=B2]| \theta'_{A \act M} \circ \omega'_{A,M}
    		\arrow[Rightarrow,from=A1,to=B1,"{\rm id}_{{\rm id}_A}\actb \vartheta_M"]
    		\arrow[Rightarrow,from=B1,to=B2,"\Theta'_{A,M}"]
    		\arrow[Rightarrow,from=A1,to=A2,"\Theta_{A,M}"]
    		\arrow[Rightarrow,from=A2,to=B2,"\vartheta_{A \act M}"']
    	\end{tikzcd}
    \end{align*}
    commutes for all $A \in \Ob(\mc C)$ and $M \in \Ob(\mc M)$.
\end{definition}

\noindent
Given a monoidal 2-category $\mc C$  and a pair $(\mc M,\mc N)$ of $\mc C$-module 2-categories, we shall refer in the following to $\msf{2Fun}_{\mc C}(\mc M,\mc N)$ as the 2-category whose objects are $\mc C$-module 2-functors, 1-morphisms are $\mc C$-module 2-natural transformations, and 2-morphisms are $\mc C$-module modifications. Physically, given a (3+1)d topological lattice model with input (spherical fusion) 2-category $\mc C$, the 2-category $\msf{2Fun}_{\mc C}(\mc M, \mc N)$ is expected to capture the boundary operators along the interface of two gapped boundaries labelled by the $\mc C$-module 2-categories $\mc M$ and $\mc N$, respectively.

\subsection{Graphical calculus}

Following \cite{barrett2012gray,douglas2018fusion}, we shall now sketch the graphical calculus for monoidal 2-categories, which readily extends to module 2-categories. The purpose of this graphical calculus is two-fold: On the one hand, it will allow us to define gauge models of topological phases as equivalence classes of membrane-net configurations. On the other hand, it conveniently provides graphical definitions for tensors that span membrane-net state spaces.  Given a monoidal 2-category $\mc C$, objects are associated with planar regions of two-dimensional membranes, 1-morphisms are depicted as one-dimensional strings at the interfaces of regions, and 2-morphisms are represented as nodes at the intersections of strings. For instance, the three surface diagrams
\begin{equation*}
	\includeTikz{0}{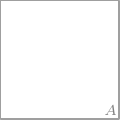}{\calculus{1}}   
	\q\q  
	\includeTikz{0}{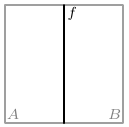}{\calculus{2}}
	\q\q 
	\includeTikz{0}{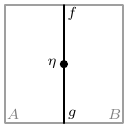}{\calculus{3}} 
\end{equation*}
are interpreted as an object $A \in \Ob(\mc C)$, a 1-morphism $f \in \Ob(\HomC_\mc C(A,B))$ and a 2-morphism $\eta \in \Hom_{\HomC_\mc C(A,B)}(f,g)$, respectively. As per our conventions, composition of 1-morphisms is depicted from left to right and vertical composition of 2-morphisms is depicted from top to bottom. This graphical calculus can then be used to depict some of the structural 2-isomorphisms entering the definitions of  the notions introduced above, e.g.
\begin{equation*}
	\includeTikz{0}{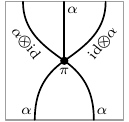}{\flatPentagonator{1}} 
	\q\q 
	\includeTikz{0}{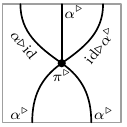}{\flatPentagonator{2}} 
	\q\q 
	\includeTikz{0}{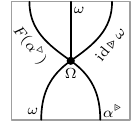}{\flatPentagonator{3}} \, ,
\end{equation*}
where we omitted the objects for convenience. Similarly, the corresponding coherence relations written earlier as pastings of commutative diagrams can  be reproduced using this graphical calculus. As a rule, we can omit to label membranes, string and nodes associated with identity object, 1- and 2-morphisms, respectively.

Thinking of the previous diagrams as being located in unit 2-cubes $[0,1]^2$, the monoidal structure can be explicitly accounted for by embedding several such diagrams in a unit 3-cube $[0,1]^3$ in such a way that the superimposition of membranes and strings represent their monoidal products. For instance, the four surface diagrams
\begin{equation*}
	\includeTikz{0}{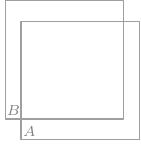}{\monoCalculus{1}}
	\q\q 
	\includeTikz{0}{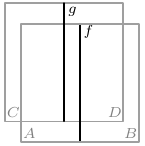}{\monoCalculus{2}} 
	\q\q 
	\includeTikz{0}{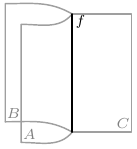}{\monoCalculus{3}} \q\q 
	\includeTikz{0}{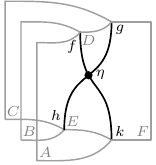}{\associator} 
\end{equation*}
shall be interpreted  as the object $A \otimes B \in \Ob(\mc C)$, the 1-morphism $f \otimes g \in \Ob(\HomC_\mc C(A \otimes C, B \otimes D))$, the 1-morphism $f \in \Ob(\HomC_\mc C(A \otimes B, C))$, and the 2-morphism $\eta \in \Hom_{\HomC(\mc C)}((f \otimes {\rm id}_C) \circ g , ({\rm id}_A \otimes h) \circ k)$, respectively. The same graphical calculus can be employed for module 2-categories. In particular, we can represent the action 2-functor $\act$ the same way we depict the monoidal product $\otimes$ by superimposing the corresponding membranes and strings. We shall clarify that a given diagram represents the module structure as opposed to the monoidal structure by using a distinct notation for the objects and/or 1-morphisms in the module 2-category. 

In prevision for the following constructions, let us now specialise to the monoidal 2-category $\VG$ of $G$-graded 2-vector spaces presented in ex.~\ref{ex:2vec}. Recall that the 2-category $\VG$ has $|G|$-many simple objects denoted by $\Vect_{g \in G}$ and its monoidal product is defined on simple objects as $\Vect_g \boxtimes \Vect_h \cong \Vect_{gh}$, for all $g,h \in G$. Moreover, the only non-trivial structural 1- or 2-isomorphism is the pentagonator, which is characterised by the group 4-cocycle $\pi$. As alluded to earlier, the 2-category $\VG$ is endowed with additional structures and satisfies many more properties than those defining a (linear) monoidal 2-category so that it defines a  \emph{spherical fusion 2-category} in the sense of Douglas and Reutter \cite{douglas2018fusion}. As a fusion 2-category, it is in particular \emph{finite} and \emph{semi-simple}---still in the sense of \cite{douglas2018fusion}. Moreover, the monoidal unit is simple and every object has left and right \emph{duals}. Specifically, given an object $\msf V \in \Ob(\VG)$, left and right duals agree and are given by the object $\msf V^\vee = \bigboxplus_{g \in G} (\msf V^\vee)_g$ with $(\msf V^\vee)_g = \HomC_{\TVect}(\msf V_{g^{-1}},\Vect)$. It follows that the dual of a simple object $\Vect_g$ is given by $\Vect_{g^{-1}}$. The sphericality further supposes the existence of \emph{pivotal} structures for objects and 1-morphisms ensuring amongst other things that membranes can be `folded' and strings can `move' within membranes. In particular, given $\Vect_g, \Vect_h, \Vect_k \in \Ob(\VG)$, the pivotal structure induces canonical equivalences of the form $\HomC_{\VG}(\Vect_g \boxtimes \Vect_h , \Vect_k)  \cong \HomC_{\VG}(\mathbb 1, \Vect_g \boxtimes \Vect_h \boxtimes \Vect_{k^{-1}})$
implying that $\HomC_{\VG}(\mathbb 1, \Vect_g \boxtimes \Vect_h \boxtimes \Vect_{k^{-1}})$ only depends on the cyclic order of the objects. More generally, it is expected that the 2-morphism in $\VG$ represented by a given surface diagram is invariant under \emph{isotopy} \cite{barrett2012gray,douglas2018fusion}. By definition, the hom-category $\HomC_{\VG}(\mathbb 1, \Vect_g \boxtimes \Vect_h \boxtimes \Vect_{k^{-1}})$ is the \emph{terminal} category unless $k = gh$. Henceforth, we thus label every membrane regions by simple objects such that the previous condition, and adaptation thereof, holds at every string. Every hom-category is thus equivalent to $\HomC_{\VG}(\mathbb 1, \mathbb 1) \cong \Vect$ and we label every string by the unique simple object in $\Vect$, namely $\mathbb C$. It follows that 2-morphisms take value in hom-sets isomorphic to $\mathbb C$ and we label every node by an arbitrary basis vector in them.  Instead of explicitly labelling every such strings and nodes, we shall make use of graphical shorthands, namely  $\includeTikz{0}{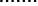}{\speString}$ and $\includeTikz{0}{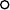}{\speNode}$.

Let us finally describe a couple of useful graphical identities. Given $g,h,k,l \in G$, the component $\pi_{\Vect_g, \Vect_h, \Vect_k, \Vect_l}$ of the invertible modification $\pi$ is a linear isomorphism whose unique non-vanishing matrix entry is $\pi(g,h,k,l) \in \mathbb C$ such that
\begin{equation}
	\label{eq:pentagonator}
	\includeTikz{0}{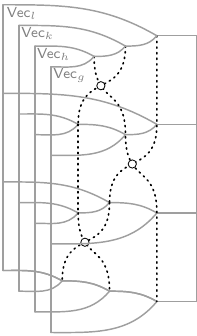}{\pentagonator{1}} = \pi(g,h,k,l) \;  
	\includeTikz{0}{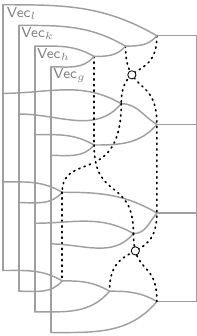}{\pentagonator{2}} \, ,
\end{equation}
i.e. the vector associated with the surface diagram depicted on the l.h.s. corresponds to the same 2-morphism in $\VG$ as that associated with the surface diagram on the r.h.s. times $\pi(g,h,k,l)$. Furthermore, given two 1-morphisms $f,g \in \HomC_\mc C(A,B)$ in a spherical fusion 2-category $\mc C$, there is a non-degenerate \emph{pairing} $\Hom_\mc C(f,g) \otimes \Hom_\mc C(g,f) \to \mathbb C$, which, together with its canonical \emph{copairing}, ensures that the vector spaces $\Hom_\mc C(f,g)$ and $\Hom_\mc C(g,f)$ are dual to one another \cite{douglas2018fusion}. Choosing adequately the normalisation of basis vectors, it gives rise in the context of $\VG$ to graphical relations of the form
\begin{equation}
	\label{eq:innerProd}
	\includeTikz{0}{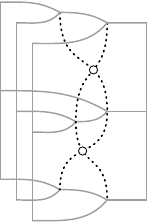}{\stack{1}} 
	\;  = \; 
	\includeTikz{0}{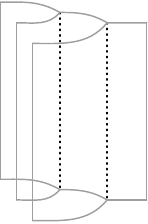}{\stack{2}} 
	\q {\rm and} \q 
	\includeTikz{0}{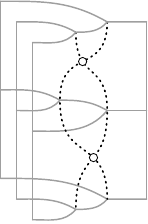}{\stack{3}}
	\; = \; 
	\includeTikz{0}{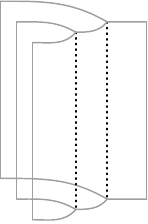}{\stack{4}}    \, .
\end{equation}

\subsection{Membrane-net state spaces}

Given a finite group $G$, the Dijkgraaf-Witten construction yields a state-sum invariant $\mc Z_G^\pi$ of 4-manifolds that has a lattice gauge theory construction \cite{dijkgraaf1990topological}. This invariant agrees with that defined by Douglas and Reutter in \cite{douglas2018fusion} for the spherical fusion 2-category $\VG$. We are interested in the vector space this state-sum invariant assigns to a given closed oriented 3-manifold. This vector space can be conveniently found as the ground state subspace of the lattice Hamiltonian realisation, where the local operators are obtained by evaluating the partition function on a special class of \emph{pinched interval cobordisms} \cite{Bullivant:2019fmk}. Nevertheless, we shall prefer at first a description whereby state spaces are defined as equivalence classes of \emph{membrane-net} configurations in the spirit of \cite{alex2011stringnet}. This alternative description, which relies on the previous graphical calculus, is more closely related to the tensor network approach presented in the following section.

Let $\Sigma$ be a closed oriented three-dimensional manifold endowed with a triangulation $\Sigma_\triangle$. We equip the triangulation $\Sigma_\triangle$ with a global ordering $v_0 < v_1 < \ldots$ of its vertices, providing a choice of \emph{branching structure}. This choice of branching structure further assigns an orientation $\epsilon(\triangle^{(n)})= \pm 1$ to every $n$-simplex $\triangle^{(n)} = (v_0v_1 \ldots v_n)$ of $\Sigma_\triangle$. We consider the \emph{polyhedral decomposition} $\Sigma_{\pentsss}$ dual to $\Sigma_\triangle$, such that every $n$-cell $\pentagon^{(n)}$ of $\Sigma_{\pentsss}$ inherits an orientation $\epsilon(\pentagon^{(n)}) = \pm 1$ from that of the dual ($3$$-$$n$)-simplex $\triangle^{(3-n)}$. For convenience, we shall label the $n$-cells of $\Sigma_{\pentsss}$ by $(v_0 v_1 \ldots v_{3-n})$ in reference to the dual ($3$$-$$n$)-simplices.

We define a $\VG$-colouring of $\Sigma_{\pentsss}$ as a map $\fr g$ that assigns a simple object $\fr g(v_0v_1) \in \Ob(\VG)$ to every oriented 2-cell $\pentagon^{(2)} \equiv (v_0v_1) \subset \Sigma_{\pentsss}$ such that $\fr g(v_0v_1) \boxtimes \fr g(v_1v_2) \cong \fr g(v_0v_2)$ for every 1-cell $\pentagon^{(1)} \equiv (v_0v_1v_2) \subset \Sigma_{\pentsss}$, the unique simple 1-morphism $\fr g(v_0v_1v_2) \simeq \mathbb C$ in the hom-category $\HomC_{\VG}(\fr g(v_0v_1) \boxtimes \fr g(v_1v_2), \fr g(v_0v_2)) \cong \Vect$ to every oriented 1-cell $\pentagon^{(1)} \equiv (v_0v_1v_2) \subset \Sigma_{\pentsss}$, and the unique basis vector $\fr g(v_0v_1v_2v_3)$ in the vector space $V^{\epsilon(\pentsss^{(0)})}[\pentagon^{(0)},\fr g]$ to every oriented 0-cell $\pentagon^{(0)} \equiv (v_0v_1v_2v_3) \subset \Sigma_{\pentsss}$, where\footnote{It follows from previous considerations that for any $\pentagon^{(0)}$, the vector spaces $V^+[\pentagon^{(0)}, \fr g] $ and $V^-[\pentagon^{(0)}, \fr g] $ are dual to one another, giving rise to graphical identities of the form \eqref{eq:innerProd}.}
\begin{align*}
	V^+[\pentagon^{(0)}, \fr g] 
	&:= \Hom_{\HomC (\mc C)}\! \big((\fr g(v_0v_1v_2)\boxtimes {\rm id}_{\fr g(v_2v_3)} ) \circ \fr g(v_0v_2v_3), ({\rm id}_{\fr g(v_0v_1)} \boxtimes \fr g(v_1v_2v_3)) \circ \fr g(v_0v_1v_3)\big)
	\\
	V^-[\pentagon^{(0)},\fr g] 
	&:= \Hom_{\HomC (\mc C)}\! \big(({\rm id}_{\fr g(v_0v_1)} \boxtimes \fr g(v_1v_2v_3)) \circ \fr g(v_0v_1v_3), (\fr g(v_0v_1v_2)\boxtimes {\rm id}_{\fr g(v_2v_3)} ) \circ \fr g(v_0v_2v_3)\big) \, .
\end{align*}
We write the resulting $\VG$-coloured decomposition as $\fr g(\Sigma_{\pentsss})$. Given a decomposition $\Sigma_{\pentsss}$, the collection of all possible $\VG$-colourings is denoted by ${\rm Col}[\Sigma_{\pentsss}, \VG]$. Finally, we consider the vector space $\mc V[\Sigma,\VG]$ of finite formal $\mathbb C$-linear combinations of $\VG$-coloured polyhedral decompositions of $\Sigma$ defined in this way.

Given $\Sigma_{\pentsss}$ and $\fr g(\Sigma_{\pentsss}) \in \mc V[\Sigma,\VG]$, let us consider an embedded 3-disk $\mathbb D^3 \hookrightarrow \Sigma$ such that $\Sigma_{\pentsss}$ is transversal to $\partial \mathbb D^3$, i.e. there are no 0-cells $\pentagon^{(0)}\subset \Sigma_{\pentsss}$ on $\partial \mathbb D^3$ and the higher-dimensional cells go through $\partial \mathbb D^3$. In virtue of the graphical calculus of $\VG$, the restriction $\fr g(\Sigma_{\pentsss}){\sss |}_{\Sigma_{\pentssss} \cap\mathbb D^3}$ of the $\VG$-coloured polyhedral decomposition $\fr g(\Sigma_{\pentsss})$ to $\Sigma_{\pentsss} \cap \mathbb D^3$ can be interpreted as the surface diagram for a vector in $\bigotimes_{\pentsss^{(0)} \subset \Sigma_{\pentssss} \cap \mathbb D^3} V^{\epsilon(\pentsss^{(0)})}[\pentagon^{(0)}, \fr g]$, obtained as the composition of all the 2-morphisms in $\VG$ associated with the 0-cells in $\Sigma_{\pentsss} \cap \mathbb D^3$. Let us now consider another $\VG$-colouring $\fr h(\Sigma_{\pentsss'})$ of a polyhedral decomposition $\Sigma_{\pentsss'}$ such that $\fr g(\Sigma_{\pentsss}){\sss |}_{\Sigma_{\pentssss} \backslash \mathbb D^3} = \fr h(\Sigma_{\pentsss'}){\sss |}_{\Sigma_{\pentssss'} \backslash \mathbb D^3}$ and the 2-morphism  associated with $\fr h(\Sigma_{\pentsss'}){\sss |}_{\Sigma_{\pentssss'}\cap \mathbb D^3}$ is the same as that associated with $\fr g(\Sigma_{\pentsss}){\sss |}_{\Sigma_{\pentssss}\cap \mathbb D^3}$ as per the graphical calculus of $\VG$. We would like to identify $\fr g(\Sigma_{\pentsss})$ and $\fr h(\Sigma_{\pentsss'})$ in $\mc V[\Sigma,\VG]$ and more generally any two $\VG$-coloured decompositions of $\Sigma$ obtained in this way. This identification can be formalised via the introduction of \emph{null} polyhedral decompositions of $\Sigma$. Given an embedded $3$-disk $\mathbb D^3 \hookrightarrow \Sigma$, we define a $\VG$-null decomposition relative to $\mathbb D^3$ as a state $\fr n(\Sigma, \mathbb D^3)$ in $\mc V[\Sigma,\VG]$ of the form $\fr n_1(\Sigma_{\pentsss_1}) + \fr n_2(\Sigma_{\pentsss_2}) + \cdots + \fr n_n(\Sigma_{\pentsss_n})$ such that every $\Sigma_{\pentsss_i}$ is transversal to $\partial \mathbb D^3$, $\fr n_1 (\Sigma_{\pentsss_1}){\sss |}_{\Sigma_{\pentssss_1}\backslash \mathbb D^3} = \ldots = \fr n_n(\Sigma_{\pentsss_n})_{\Sigma_{\pentssss_n}\backslash \mathbb D^3}$, and $\fr n_1(\Sigma_{\pentsss_1}){\sss |}_{\Sigma_{\pentssss_1}\cap \mathbb D^3} + \cdots + \fr n_n(\Sigma_{\pentsss_n}){\sss |}_{\Sigma_{\pentssss_n}\cap \mathbb D^3} = 0$ as a 2-morphism in $\VG$. An explicit example of such a $\VG$-null polyhedral decomposition is provided by $\fr n_1(\Sigma_{\pentsss_1}) - \fr n_2(\Sigma_{\pentsss_2})$ where $\fr n_1(\Sigma_{\pentsss_1}){\sss |}_{\Sigma_{\pentssss_1}\cap \mathbb D^3}$ and  $\fr n_2(\Sigma_{\pentsss_2}){\sss |}_{\Sigma_{\pentssss_2}\cap \mathbb D^3}$ are depicted on the l.h.s. and r.h.s. on eq.~\eqref{eq:pentagonator}, respectively.\footnote{This $\VG$-null decomposition can be thought as a $2 \leftrightharpoons 3$ \emph{Pachner move} on the dual triangulation. In general, every $\VG$-null decomposition relative to a 3-disk corresponds to a sequence of three-dimensional Pachner moves \cite{PACHNER1991129}.} We can then define the subspace $\mc V_{\varnothing}[\Sigma,\VG] \subset \mc V[\Sigma,\VG]$ of finite formal linear combinations of null $\VG$-coloured polyhedral decompositions of $\Sigma$ for any collection of embedded $3$-disks $\mathbb D^3 \hookrightarrow \Sigma$.
Putting everything together, we define the \emph{membrane-net} state space associated with $\Sigma$ as the quotient space
\begin{equation}
	\label{eq:membSpace}
	\mc M[\Sigma, \VG]:= \mc V[\Sigma, \VG] \, \big/ \, \mc V_{\varnothing}[\Sigma, \VG] \, .
\end{equation}
As we mentioned earlier, there exist alternative constructions of the membrane-net space $\mc M[\Sigma,\VG]$. In particular, it is isomorphic to the ground state subspace of the Hamiltonian realisation of Dijkgraaf-Witten theory. We shall briefly sketch here this isomorphism mimicking Kirillov's treatment of \emph{string-net} spaces \cite{alex2011stringnet}. Given a triangulation $\Sigma_{\triangle}$ of $\Sigma$ and its dual polyhedral decomposition $\Sigma_{\pentsss}$, we consider the state space
\begin{equation}
	\label{eq:Hspace}
	\mc H_G^\pi[\Sigma_\triangle]= \bigoplus_{\fr g \in {\rm Col}[\Sigma_\triangle,\VG]} \bigotimes_{\triangle^{(3)}\subset \Sigma_{\triangle}}V^{\epsilon(\triangle^{(3)})}[\triangle^{(3)}, \fr g] \, .
\end{equation}
It follows from the \emph{boundary relative triangulation independence} and the \emph{unitarity} condition of the Dijkgraaf-Witten state-sum $\mc Z^\pi_G$ that the linear map $\mc Z^\pi_G(\Sigma_{\triangle}\times \mathbb I) : \mc H^\pi_G[\Sigma_{\triangle}] \to \mc H^\pi_G[\Sigma_{\triangle}]$ is an \emph{Hermitian projector} that coincides with the ground state projector of the lattice Hamiltonian realisation of the theory on $\Sigma_{\triangle}$. Therefore, the quantum invariant assigned to $\Sigma_{\triangle}$ is the vector space $\mc Z_G^\pi(\Sigma_{\triangle}):= {\rm Im}\, \mc Z_G^\pi(\Sigma_{\triangle}\times \mathbb I)$. The boundary relative triangulation independence further ensures that given two triangulations $\Sigma_{\triangle}$ and $\Sigma_{\triangle'}$ of $\Sigma$, the vector spaces $\mc Z^\pi_G(\Sigma_{\triangle})$ and $\mc Z^\pi_G(\Sigma_{\triangle'})$ are isomorphic \cite{Bullivant:2019fmk}. We thus notate the quantum invariant assigned to $\Sigma$ via $\mc Z^\pi_G(\Sigma)$. Our task amounts to showing that $\mc M[\Sigma,\VG] \simeq \mc Z^\pi_G(\Sigma)$. 

Let us consider the membrane-net space $\mc M[\Sigma \backslash \Sigma_{\triangle}^{(0)}, \VG]$, where $\Sigma_\triangle^{(0)}$ is the 0-skeleton of the triangulation $\Sigma_\triangle$. Since removing points from $\Sigma$ prevents the membranes of the surface diagram canonically associated with $\Sigma_{\pentsss}$ to move freely, and thus constrains the kind of graphical relations encoded into the states in $\mc V_{\varnothing}[\Sigma,\VG]$, it follows from  the definition of membrane-net state spaces that $\mc M[\Sigma \backslash \Sigma_{\triangle}^{(0)}, \VG] \simeq \mc H^\pi_G[\Sigma_\triangle]$. Crucially, there is a trick to restore the rules of the graphical calculus that consists in inserting the normalised sum of all possible $\VG$-colourings of a spherical membrane around every point in $\Sigma_{\triangle}^{(0)}$. It can  be  readily checked using the graphical calculus away from $\Sigma^{(0)}_{\triangle}$ that these insertions have the effect of making the defects left by the points in the 0-skeleton of $\Sigma_{\triangle}$ invisible to the membranes/2-cells. Denoting by $\mathbb A_{\triangle^{(0)}} : \mc M[\Sigma \backslash \Sigma_{\triangle}^{(0)}, \VG] \to \mc M[\Sigma \backslash \Sigma_{\triangle}^{(0)}, \VG]$ the local projector that performs such an insertion around the defect left by the 0-simplex $\triangle^{(0)}$, we have an isomorphism ${\rm Im} \big( \prod_{\triangle^{(0)}\subset \Sigma_\triangle}\mathbb A_{\triangle^{(0)}} \big)\simeq \mc M[\Sigma,\VG]$. By expressing these local projectors in terms of what the partition function assigns to certain pinched interval cobordisms as in \cite{Bullivant:2019fmk}, we can check that $\prod_{\triangle^{(0)}\subset \Sigma_\triangle}\mathbb A_{\triangle^{(0)}} = \mc Z_G^\pi(\Sigma_{\triangle}\times \mathbb I)$. Putting everything together, we obtain $\mc M[\Sigma,\VG] \simeq \mc Z^\pi_G(\Sigma)$.

\section{Topological tensor network states\label{sec:TN}}

\emph{In this section, we define tensor network representations of the membrane-net state spaces introduced previously, derive the corresponding virtual symmetries, and relate the existence of two canonical tensor network representations to the electromagnetic duality of the model. Finally, we specialise to the three-dimensional toric code.}

\subsection{Tensor network representations}

Given a closed oriented three-dimensional surface $\Sigma$, we are interested in (exact) tensor network representations of the membrane-net state space $\mc M[\Sigma, \VG]$, i.e. we wish to write all possible states in $\mc M[\Sigma, \VG]$ as tensor networks. More specifically, we would like to study the possibility of defining several `inequivalent' such tensor network representations in a sense that will be made precise. Ultimately, our main focus is the relation between the existence of two canonical tensor network representations of $\mc M[\Sigma, \TVect_G]$ and the electromagnetic duality of the corresponding topological phase. As such, we assume in the following that the input 4-cocycle is trivial so that the pentagonator evaluates to the identity 2-isomorphism, although our construction applies to the general case as well.

As we mentioned in the introduction, the tensors we consider  carry both virtual indices labelling entanglement degrees of freedom, along which tensors are contracted to one another, and physical indices, which are left uncontracted. The individual physical vector spaces are chosen so that their tensor product corresponds to the state space of a concrete physical system. We shall restrict for now our attention on tensors that possess one set of physical indices and four sets of virtual indices. Endowing each of these tensors with the geometry of a 3-simplex, let us suppose that contraction between two tensors take place when they share a 2-simplex. The graph underlying a tensor network built up in this way can thus be identified with a delta-complex. Conversely, given a triangulation $\Sigma_{\triangle}$ of $\Sigma$, we can assign such a tensor to every 3-simplex $\triangle^{(3)} \subset \Sigma_{\triangle}$. We then require the physical space of the resulting tensor network to be isomorphic to $\mc H_G[\Sigma_{\triangle}]$ as defined in eq.~\eqref{eq:Hspace}, which in turn dictates what the individual physical spaces should be. In other words, the tensor network is a wave function in a vector space that is governed by the membrane-net state space we are trying to describe, and more specifically by the input spherical 2-category $\TVect_G$. Nevertheless, given the ancillary nature of the virtual indices, there is some freedom regarding the vector spaces they can belong to. This is this freedom that we want to exploit.

We would like to argue that a family of tensor network representations of $\mc M[\Sigma,\TVect_G]$ are indexed by finite semi-simple module 2-categories over $\TVect_G$.  Given a finite semi-simple left $\TVect_G$-module 2-category $\mc M$, we consider a tensor $\mc T^{\mc M}_{\TVect_G}$ with one set of physical indices and four sets of virtual indices that is best defined graphically via the middle diagram below:
\begin{equation}
    \label{eq:simplex}
    \includeTikz{0}{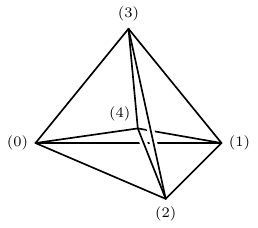}{\fourSimplex{1}} \!\!\!\!\!
    \longleftrightarrow \;
    \mc T^{\mc M}_{\TVect_G}  
    \equiv 
    \includeTikz{0}{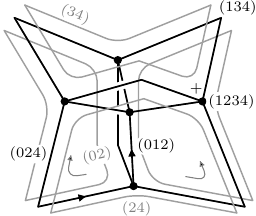}{\simplex{1}{}{}{}{}}
    \;\;\equiv \;\;
    \includeTikz{0}{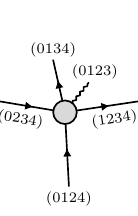}{\convTensor} \, ,
\end{equation}
where every membrane, string and node of the drawing is associated with an index of the tensor. The various constituents of this diagram are oriented according to the branching structure of the `dual' 4-simplex $(\snum{01234})$ depicted on the left-hand side, which is positively oriented by convention. We explicitly indicated this correspondence in a few cases, the remaining ones can be deduced by analogy. We qualify all the membranes, strings and nodes identified with simplices containing the 0-simplex $(\snum{4})$ as `virtual' and the other ones as `physical'. The indices of $\mc T^{\mc M}_{\TVect_G}$ inherits this terminology. Combining together via a tensor product all the indices associated with one of the five 3-simplices in the boundary of the 4-simplex yields the more conventional notation depicted on the r.h.s. of eq.~\eqref{eq:simplex}, where the squiggly line represents the combined physical index. We should then think of the combined virtual indices as being dual to the faces of the 3-simplex $(\snum{0123})$. 

Our graphical depiction of the tensor $\mc T^{\mc M}_{\TVect_G}$ is borrowed from the graphical calculus of  2-categories in such a way that indices associated with membranes, string and nodes are labelled by objects, 1-morphisms and 2-morphisms, respectively, in a given 2-category.\footnote{Note that for visual clarity we had to perform one contortion to the graphical calculus of monoidal and module 2-categories in such a way that the membranes identified with 1-simplices $(v_0\ftnum{4})$ and $(v_1\ftnum{4})$, respectively, should be thought as meeting the membrane $(v_0v_1)$ along the string identified with $(v_0v_1\ftnum{4})$.} More precisely, we label physical membranes identified with 1-simplices $(v_0v_1)$ by simple objects $\fr g(v_0v_1) \in \Ob(\TVect_G)$, virtual membranes $(v_0 \snum{4})$ by simple objects $\fr m(v_0 \snum{4}) \in \sOb(\mc M)$, physical strings $(v_0v_1v_2)$ by the unique simple 1-morphism $\fr g(v_0v_1v_2)$ in the hom-category $\HomC_{\TVect_G}(\fr g(v_0v_1)\boxtimes \fr g(v_1v_2), \fr g(v_0v_2))$, virtual strings $(v_0v_1\snum{4})$ by simple 1-morphisms $\fr m(v_0v_1\snum{4}) \in \sOb(\HomC_{\mc M}(\fr g(v_0v_1) \act \fr m(v_1\snum{4}), \fr m(v_0\snum{4})))$, the physical node $(\snum{0123})$ by the unique basis vector $\fr g(\snum{0123})$ in the vector space $V^+[(\snum{0123}), \fr g]$, and finally virtual nodes $(v_0v_1v_2\snum{4})$ by basis vectors $\fr m(v_0v_1v_2\snum{4})$ in the vector spaces $V^{\epsilon(v_0v_1v_2\ftnum{4})}[(v_0v_1v_2\snum{4}),\fr g, \fr m]$ given by
\begin{align*}
    V^+[(v_0v_1v_2\snum{4}), \fr g, \fr m] 
    &= \Hom_{\HomC(\mc M)}\! \big((\fr g(v_0v_1v_2) \act {\rm id}_{\fr m(v_2\ftnum{4})} ) \circ \fr m(v_0v_2\snum{4}), 
    ({\rm id}_{\fr g(v_0v_1)} \act \fr m(v_1v_2\snum{4})) \circ \fr m(v_0v_1\snum{4})\big)
    \\
    V^-[(v_0v_1v_2\snum{4}),\fr g, \fr m] 
    &= \Hom_{\HomC (\mc M)}\! \big(({\rm id}_{\fr g(v_0v_1)} \act \fr m(v_1v_2\snum{4})) \circ \fr m(v_0v_1\snum{4}),
    (\fr g(v_0v_1v_2)\act {\rm id}_{\fr m(v_2\ftnum{4})} ) \circ \fr m(v_0v_2\snum{4})\big) .
\end{align*}
In the above $\epsilon(v_0v_1v_2\snum{4})$ equals $+1$ if the orientation of $(v_0v_1v_2\snum{4})$ induced by the branching structure agrees with that induced by the orientation of the 4-simplex $(\snum{01234})$, and $-1$ otherwise.
Non-vanishing components of $\mc T^{\mc M}_{\TVect_G}$ require every hom-category to be non-terminal and every vector space of 2-morphisms to be non-trivial. In particular, we established that hom-categories associated with physical strings $(v_0v_1v_2)$ are terminal unless $\fr g(v_0v_1) \boxtimes \fr g(v_1v_2) \cong \fr g(v_0v_2)$, in which case $\HomC_{\TVect_G}(\fr g(v_0v_1)\boxtimes \fr g(v_1v_2), \fr g(v_0v_2)) \cong \Vect$ and the unique simple 1-morphism is $\fr g(v_0v_1v_2)$ is isomorphic to $\mathbb C$. The non-vanishing components of $\mc T^{\mc M}_{\TVect_G}$ finally evaluate to the matrix entries of maps
\begin{equation}
    V^+[(\snum{1234}), \fr g, \fr m] 
    \otimes
    V^+[(\snum{0134}), \fr g, \fr m] 
    \otimes
    V^+[(\snum{0123}), \fr g, \fr m] 
    \xrightarrow{\sim}     
    V^+[(\snum{0234}), \fr g, \fr m] 
    \otimes
    V^+[(\snum{0124}), \fr g, \fr m]  \, ,
\end{equation}
the collection of which specifies the module pentagonator components $\pi^\act_{\fr g(01),\fr g(12), \fr g(23), \fr m(34)}$ in $\mc M$. In case the 4-simplex $(\snum{01234})$ were to have \emph{opposite} orientation,  we would define the non-vanishing components to evaluate to the matrix elements of the \emph{inverse} of the module pentagonator. Practically, given a branched triangulation $\Sigma_{\triangle}$ of a three-dimensional surface $\Sigma$, a state in $\mc M[\Sigma,\TVect_G]$ is obtained by assigning to every $3$-simplex $(v_0v_1v_2v_3) \subset \Sigma_\triangle$ a copy of the tensor $\mc T^\mc M_{\TVect_G}$ following the conventions spelt out in this section for the 4-simplex $(v_0v_1v_2v_3v_3')$ such that $v_3 < v_3' < v_4$. Instead of investigating the case of a general left $\TVect_G$-module 2-category $\mc M$ in detail---this would require delving deeper into the theory of spherical fusion 2-categories and module 2-categories---we shall immediately specialise to two tensor network representations associated with canonical choices of $\mc M$, and then infer results for the general case. 

The first representation we wish to focus on is obtained by choosing the module 2-category $\TVect_G$ over itself. The virtual membranes of the tensor $\mc T^{\TVect_G}_{\TVect_G}$ are thus labelled by simple objects in $\TVect_G$. Analogously to the physical hom-categories, the virtual hom-categories are non-terminal unless $\fr g(v_0v_1) \act \fr m(v_1v_2) \cong \fr m(v_0v_2)$ for every virtual string identified with $(v_0v_1v_2)$, in which case they are all equivalent to $\Vect$. Therefore, non-vanishing components of $\mc T^{\TVect_G}_{\TVect_G}$ are such that all the strings are labelled by a simple 1-morphism that is isomorphic to the unique simple object in $\Vect$, namely $\mathbb C$, and all the nodes are labelled by the unique basis vector in the vector spaces of 2-morphisms, which are all isomorphic to $\mathbb C$. By definition, the module pentagonator is equal to the monoidal pentagonator in $\TVect_G$, which evaluates to the identity 2-isomorphism. Exploiting our notations for surface diagrams in $\TVect_G$, the non-vanishing components of $\mc T^{\TVect_G}_{\TVect_G}$ thus read
\begin{equation}
	\label{eq:simplexVecG}
	\includeTikz{0}{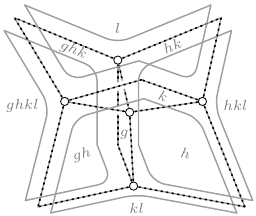}{\simplex{2}{g}{h}{k}{l}} = 1 \, , \q \forall \, g,h,k,l \in G \, ,
\end{equation}
where we identified simple objects $\Vect_g \in \Ob(\TVect_G)$ and the corresponding group variables $g \in G$ for visual clarity.
In order to confirm that the tensor $\mc T^{\TVect_G}_{\TVect_G}$ does yield a representation of the membrane-net space $\mc M[\Sigma, \TVect_G]$, it is convenient to employ the alternative definition presented at the end of the previous section. Given a branched triangulation $\Sigma_\triangle$ of $\Sigma$, we assign one tensor to every oriented 3-simplex as explained above. Notice that when contracting two tensors, in addition to virtual indices being summed over, the indices associated with the physical strings along which the contraction takes place are identified. Similarly, when contracting several such tensors around a 1-simplex of $\Sigma_\triangle$, the indices associated with the physical membranes along which the contraction takes place are identified. 
It follows that the resulting tensor network state is valued in a vector space isomorphic to $\mc H_G[\Sigma_\triangle]$. As a matter of fact, this statement applies for every tensor network representation we consider. It remains to confirm that the tensor network state is in the image of the ground state projector $\prod_{\triangle^{(0)}\subset \Sigma_\triangle}\mathbb A_{\triangle^{(0)}}$. Firstly, notice that given the geometry of $\mc T^{\TVect_G}_{\TVect_G}$, the virtual membranes around every $\Sigma_{\triangle^{(0)}} \subset \Sigma_\triangle$ combine so as to form a closed membrane whose topology is that of the 2-sphere, and the corresponding label is summed over in virtue of the contractions. Secondly, recall that given $\triangle^{(0)}\subset \Sigma_{\triangle}$, the operator $\mathbb A_{\triangle^{(0)}}$ amounts to inserting the normalised sum of all possible $\TVect_G$-colourings of a spherical membrane around $\triangle^{(0)}$. But it follows from the graphical calculus in $\TVect_G$ that two nested spherical membranes whose object labels are summed over is the same thing, up to a factor of $\frac{1}{|G|}$, as one spherical membrane whose label is summed over---this is nothing but the statement that $\mathbb A_{\triangle^{(0)}} \circ \mathbb A_{\triangle^{(0)}} = \mathbb A_{\triangle^{(0)}}$. It follows that tensor network states obtained in this way are in the image of $\prod_{\triangle^{(0)}\subset \Sigma_\triangle}\mathbb A_{\triangle^{(0)}}$ and thus define states in $\mc M[\Sigma,\TVect_G]$.

The second canonical representation is obtained by choosing $\mc M= \TVect$, which is a (finite semi-simple) $\TVect_G$-module 2-category via the forgetful functor $\TVect_G \to \TVect$ with trivial $G$-action 2-functor. The virtual membranes of $\mc T^{\TVect}_{\TVect_G}$ are labelled by the single simple object in the 2-category $\TVect$, namely $\Vect$, that we shall depict as a planar region with a dotted grey contour. Virtual hom-categories are all equivalent to $\Vect$, and thus we label the corresponding strings with the unique simple object $\mathbb C \in \Ob(\Vect)$. All the vector spaces of 2-morphisms are one-dimensional and the corresponding nodes are labelled by the unique basis vectors. As for the previous representation, the module pentagonator in $\TVect$ evaluates to the identity 2-morphism so that the non-vanishing components of the tensor $\mc T^{\TVect}_{\TVect_G}$ are all equal to $1 \in \mathbb C$. Graphically, these non-vanishing components read
\begin{equation}
	\label{eq:simplexVec}
	\includeTikz{0}{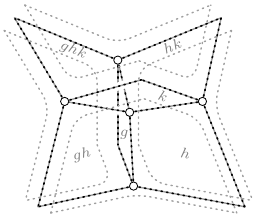}{\simplex{3}{g}{h}{k}{}} = 1 \, , \q \forall \, g,h,k \in G \, .
\end{equation}
The fact that the tensor $\mc T^{\TVect}_{\TVect_G}$ does yield a representation of the membrane-net space $\mc M[\Sigma, \TVect_G]$ follows from definition given in eq.~\eqref{eq:membSpace} and the associahedron axiom satisfied by the module pentagonator (see def.~\ref{def:mod2Cat}). Indeed, in virtue of the graphical calculus of $\TVect_G$ tensor network states obtained by contracting copies of $\mc T^{\TVect}_{\TVect_G}$ readily provides $\TVect_G$-coloured polyhedral decompositions of $\Sigma$ up to local relations encoded into the associahedron axiom. These local relations, which are associated with three-dimensional Pacher moves, are in one-to-one correspondence with the $\TVect_G$-null decompositions in $\mc V_{\varnothing}[\Sigma,\TVect_G]$. Keeping in mind that virtual indices, which take value in the module 2-category, are traced over upon contraction of the tensor network, the previous analysis largely generalises to the case of $\mc T^{\mc M}_{\TVect_G}$. In particular, the associahedron axiom of the module pentagonator $\pi^{\act}$---which is non-trivial in the case of a generic module 2-category $\mc M$---ensures that the tensor network states satisfy the local relations encoded into $\mc V_{\varnothing}[\Sigma,\TVect_G]$. As evoked earlier, a rigorous treatment of this more general case would require us to delve deeper into the theory of finite semi-simple module 2-categories, which is not the main objective of this manuscript.

\subsection{Virtual symmetry operators\label{sec:virtual}}

Topological order manifests itself in tensor network states through the existence of (non-trivial) virtual symmetry conditions, i.e. symmetries with respect to non-local operators acting solely along virtual indices \cite{PhysRevB.84.165139,PhysRevB.83.035107,SCHUCH20102153,PhysRevLett.111.090501,BUERSCHAPER2014447,Sahinoglu:2014upb,BULTINCK2017183,Bultinck_2017,Williamson:2017uzx,Delcamp:2020rds,Williamson:2020hxw,10.25365/thesis.43085}. In particular, closed versions of these operators along non-contractible cycles of the manifold may be employed to map ground states from one another. Specifically, we expect---analogously to the (2+1)d scenario---the order of a given (3+1)d topological phase to be realised as the Drinfel'd centre of the spherical fusion 2-category of virtual symmetry operators for a given tensor network representation. These symmetry conditions can be conveniently rephrased in terms of pulling-through conditions, whereby virtual operators on the entanglement level can be pulled through the tensors making them freely deformable. Given a tensor network representation $\mc T^{\mc M}_{\TVect_G}$, these virtual operators can be written in terms of a tensor $\mc O^{\mc M}_{\TVect_G}$ that does not possess any physical indices but instead five sets of virtual indices. As before, the tensor $\mc O^{\mc M}_{\TVect_G}$ is best defined graphically via the right diagram below:
\begin{equation}
    \label{eq:prism}
    \includeTikz{0}{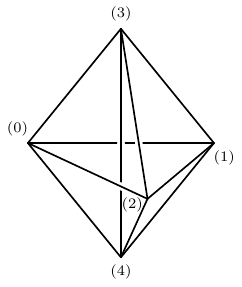}{\fourSimplex{2}} \!\!
    \longleftrightarrow \;
    \mc O^{\mc M}_{\TVect_G}  
    \equiv 
    \includeTikz{0}{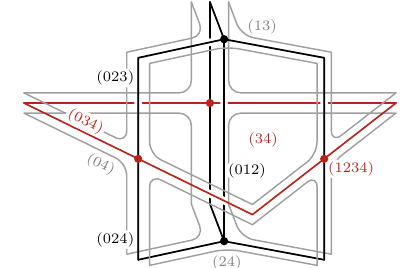}{\prism{1}{}{}{}{}{}{}{}{}{}{}{}}
    \, .
\end{equation}
Every membrane, string and node of the drawing is associated with an index of the tensor and these are oriented according to the orientations of the 1-, 2- and 3-simplices in the boundary of the `dual' 4-simplex depicted on the left following the same rules as in eq.~\eqref{eq:simplex}. In particular, the membrane delimited by the red bold line is associated with the 1-simplex $(\snum{34})$ and it does not intersect the membranes associated with the 1-simplices $(\snum{01})$, $(\snum{12})$ and $(\snum{02})$, respectively. 

Membranes, string and nodes are still labelled by objects, 1-morphisms and 2-morphisms, respectively, in a given 2-category. Some of the labelling types  are inherited from those of the tensor $\mc T^{\mc M}_{\TVect_G}$ in such a way that the two tensors can be contracted. Indeed, we label the membranes associated with the 1-simplices $(v_0v_1)$ with $v_0,v_1 = \snum{0,1,2}$ by simple objects $\fr g(v_0v_1)\in \Ob(\TVect_G)$, membranes $(v_0v_1)$ with $v_0 = \snum{0,1,2}$ and $v_1 = \snum{3,4}$ by simple objects $\fr m(v_0v_1) \in \sOb(\mc M)$, the string $(\snum{123})$ by the unique simple 1-morphism $\fr g(\snum{123})$ in the hom-category $\HomC_{\TVect_G}(\fr g(\snum{01}) \boxtimes \fr g(\snum{12}), \fr g(\snum{02}))$, strings $(v_0v_1v_2)$ with $v_0,v_1 = \snum{0,1,2}$ and $v_2=\snum{3,4}$ by simple 1-morphisms $\fr m(v_0v_1v_2) \in \sOb(\HomC_{\mc M}(\fr g(v_0v_1) \act \fr m(v_1v_2), \fr m(v_0v_2)))$, and finally nodes $(\snum{012}v_0)$ with $v_0=\snum{3,4}$ by basis vectors $\fr m(\snum{012}v_0)$ in the vector spaces $V^{\epsilon(\ftnum{012}v_0)}[(\snum{012}v_0),\fr g, \fr m]$, where $\epsilon(\snum{012}v_0)$ is the orientation of $(\snum{012}v_0)$ relative to that of the 4-simplex. Analogously to $\mc T^{\mc M}_{\TVect_G}$, non-vanishing components of $\mc O^{\mc M}_{\TVect_G}$ requires every hom-category to be non-terminal and every vector space of 2-morphisms to be non-trivial. Crucially, the previous enumeration does not contain the labellings of the indices associated with the simplices $\triangle^{(n)}$ containing $(\snum{34})$ as a subsimplex. In order to determine what the remaining seven indices should be labelled by, and what the virtual tensor should evaluate to, we must solve the pulling-through conditions encoded into the diagrammatic equation
\begin{equation}
	\label{eq:pullingThrough}
	\includeTikz{0}{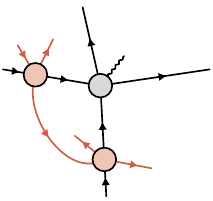}{\pullingThrough{1}} 
	\; \stackrel{!}{=} \;  
	\includeTikz{0}{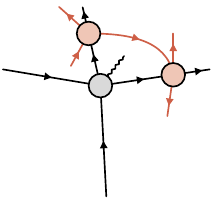}{\pullingThrough{2}} \, ,
\end{equation}
where we introduced a more conventional notation (red blob) for the tensor $\mc O^{\mc M}_{\TVect_G}$ obtained by combining all the indices associated with one of the five 3-simplices in the boundary of the corresponding 4-simplex. Recall that the tensors  $\mc T^{\mc M}_{\TVect_G}$ evaluate to the matrix elements of the components of the module pentagonator $\pi^{\act}$, or its inverse depending on the orientation of the corresponding $4$-simplex. Given that the tensors $\mc O^{\mc M}_{\TVect_G}$ have a structure very similar to that of the tensors $\mc T^{\mc M}_{\TVect_G}$, we expect them to evaluate to the matrix elements of the components of a pentagonator-like invertible modification. In this case, the pulling-through conditions would follow from an associahedron axiom involving $\pi^{\act}$.  Before commenting about the case of an arbitrary left module 2-category $\mc M$, it is best to start with our two examples. 

\medskip \noindent
Let us first consider the tensor network representation obtained by choosing $\mc M= \TVect_G$. Recall that  virtual and physical membranes of $\mc T^{\TVect_G}_{\TVect_G}$ are labelled by simple objects in $\TVect_G$ and that the tensor components vanish unless all the hom-categories are non-terminal. Similarly, all the membranes in $\mc O^{\TVect_G}_{\TVect_G}$ are labelled by simple objects in $\TVect_G$ and non-vanishing tensor components require the hom-categories to be non-terminal. Graphically, these non-vanishing components read
\begin{equation}
	\label{eq:prismVecG}
	\includeTikz{0}{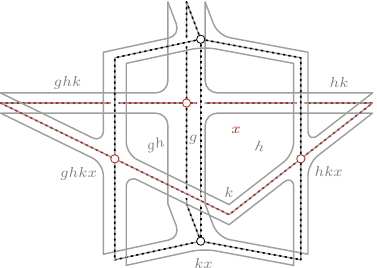}{\prism{2}{g}{h}{k}{x}{}{}{}{}{}{}{}} = 1  \, , \q \forall \, g,h,k,x \in G \, .
\end{equation}
Consider the contraction of the tensors $\mc T^{\TVect_G}_{\TVect_G}$ and $\mc O^{\TVect_G}_{\TVect_G}$ obtained by `stacking' them on top of each other as they are depicted in eq.~\eqref{eq:simplexVecG} and eq.~\eqref{eq:prismVecG}, respectively. We can then think of the membrane associated with the 1-simplex $(\snum{34})$ labelled by $\fr g(\snum{34}) \cong \Vect_x \in \Ob(\TVect_G)$ as acting from the right via the monoidal product in $\TVect_G$ on the virtual membranes of $\mc T^{\TVect_G}_{\TVect_G}$ that are contracted with the membranes $(\snum{03})$, $(\snum{13})$ and $(\snum{23})$ of $\mc O^{\TVect_G}_{\TVect_G}$. Notice that upon such a contraction, the membrane $(\snum{34})$ does not act on the physical indices of $\mc T^{\TVect_G}_{\TVect_G}$.
The pulling-through condition then readily follows from the fact that all the non-vanishing components of both $\mc T^{\TVect_G}_{\TVect_G}$ and $\mc O^{\TVect_G}_{\TVect_G}$ are equal to 1 and correspond to labellings for which all the hom-categories are non-terminal. Evidently, it follows from the monoidal structure in $\TVect_G$ that the contraction of two copies of $\mc O^{\TVect_G}_{\TVect_G}$ also fulfils the pulling-through condition in such a way that acting successively  on the virtual membranes of $\mc T^{\TVect_G}_{\TVect_G}$ with membranes labelled by $\Vect_x$ and $\Vect_y$ in $\Ob(\TVect_G)$ is identical to acting with a single membrane labelled by $\Vect_{xy}$.

The action we just described endows the left $\TVect_G$-module 2-category $\mc M = \TVect_G$ with the structure of a right $\TVect_G$-module 2-category, so that $\mc M=\TVect_G$ is equipped with the structure of a $(\TVect_G,\TVect_G)$-bimodule 2-category. The non-vanishing components of $\mc O^{\TVect_G}_{\TVect_G}$ should then be thought as the matrix elements of maps defining the bimodule left pentagonator components $\pi^{\caact}_{\Vect_g, \Vect_h, \Vect_k, \Vect_x}$ of $\TVect_G$, which evaluates to the identity 2-isomorphism. In this context, the pulling condition is ensured by the associahedron axiom involving $\pi^{\caact}$ and $\pi^{\act}$ (see def.~\ref{def:biMod2Cat}), which is trivially satisfied. 

\medskip \noindent
Let us now consider the second tensor network representation obtained by choosing $\mc M= \TVect$. Recall that virtual membranes of $\mc T^{\TVect}_{\TVect_G}$ are labelled by $\Vect$ as the unique simple object in $\TVect$, whereas physical membranes are labelled by simple objects in $\TVect_G$. As discussed earlier, the tensor $\mc O^{\TVect}_{\TVect_G}$ inherits some labellings from $\mc T^{\TVect}_{\TVect_G}$ ensuring that these can be contracted together. The remaining labellings are defined as follows: The membrane $(\snum{34})$ is labelled by a simple object $\mc N$ in $\MOD(\Vect_G^{\rm op})$, i.e. an indecomposable finite semi-simple \emph{right} module (1-)category over $\Vect_G$, the strings $(v_0\snum{34})$ with $v_0=\snum{0},\snum{1},\snum{2}$ are labelled by representatives $\fr n(v_0\snum{12})$ of isomorphism classes of simple objects in $\mc N$, and finally the three nodes $(v_0v_1\snum{34})$ are labelled by basis vectors in the vector spaces $\Hom_{\mc N}(\fr n(v_0\snum{34}) \cat \mathbb C_{(v_0v_1)}, \fr n(v_1\snum{34}))$, or their duals depending on the orientations of the 3-simplices $(v_0v_1\snum{34})$ relative to that of $(\snum{01234})$, where $\mathbb C_{(v_0v_1)} \simeq \mathbb C_{g \in G} \in \Ob(\Vect_G)$ such that $\fr g(v_0v_1)\cong \Vect_g \in \Ob(\TVect_G)$. As usual, non-vanishing elements of $\mc O^{\TVect}_{\TVect_G}$ requires every hom-category to be non-terminal and every vector space of 2-morphisms to be non-trivial. The pulling-through conditions finally impose that these non-vanishing components evaluate to the entries of the matrices defined in eq.~\eqref{eq:SixJCat} that characterise the right module associator $\alpha^\cat$ of $\mc N$. Graphically, these components read
\begin{equation}
    \label{eq:prismTVec}
    \includeTikz{0}{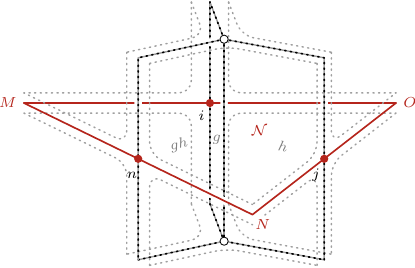}{\prism{3}{g}{h}{}{}{N}{M}{O}{n}{j}{i}{\mc N}} 
    = \SixJCat{M}{\mathbb C_g}{O}{\mathbb C_h}{N}{\mathbb C_{gh}}_{ij,1n} \, ,
\end{equation}
for any $g,h \in G$, simple objects $M,N,O \in {\rm sOb}(\mc N)$, and basis vectors $\la {}^{M\mathbb C_{gh}}_N ,n | \in \Hom_{\mc N}(M \cat \mathbb C_{gh},N)^\star$, $| {}^{M\mathbb C_g}_O,i \ra \in \Hom_{\mc N}( M \cat \mathbb C_g ,O)$, $|{}^{O \mathbb C_h}_{N},j \ra \in \Hom_{\mc N}(O \cat \mathbb C_h,N)$. The fact that the tensor $\mc O^{\TVect}_{\TVect_G}$ is indeed a solution of the pulling-through conditions illustrated in eq.~\eqref{eq:pullingThrough} follows from the labellings of the non-vanishing components of the tensor $\mc T^{\TVect}_{\TVect_G}$ provided by eq.~\eqref{eq:simplex}, together with the pentagon axiom satisfied by $\alpha^\cat$ and the monoidal associator $\alpha$ of $\Vect_G$ that happens to evaluate to the identity 1-morphism.

So we obtained that the tensor $\mc T^{\TVect}_{\TVect_G}$ satisfies symmetry conditions with respect to virtual operators labelled by simple objects in $\MOD(\Vect_G^{\rm op})$. Henceforth, we refer to this 2-category as the 2-category $\TRep(G)$ of 2-representations of $G$. By construction, the 2-category $\mc M = \TVect$ has the structure of a right module 2-category over $\TRep(G)$ via the forgetful 2-functor $\TRep(G) \to \TVect$ so that $\mc M$ is also a $(\TVect_G,\TRep(G))$-bimodule 2-category. It follows that the non-vanishing components of $\mc O^{\TVect}_{\TVect_G}$ given in eq.~\eqref{eq:prismTVec} actually correspond to the matrix elements of maps defining the bimodule left pentagonator components $\pi^{\caact}_{\Vect_g, \Vect_h, \Vect, \mc N}$ of $\TVect$ according to
\begin{equation}
	\includeTikz{0}{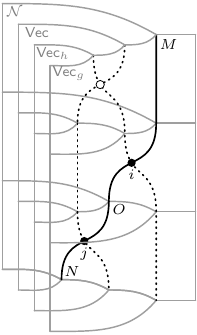}{\pentagonator{3}} = \sum_{n}
	\SixJCat{M}{\mathbb C_g}{O}{\mathbb C_h}{N}{\mathbb C_{gh}}_{ij,1n} \; 
	\includeTikz{0}{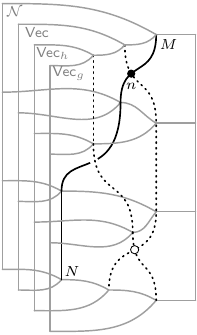}{\pentagonator{4}} 
\end{equation} 
so that the pulling-through conditions follow from the associahedron axiom involving $\pi^{\caact}$ and $\pi^{\act}$. 
Note that given our knowledge of the classification of \emph{indecomposable} module categories over $\Vect_G$ \cite{ostrik2002module}, we could have written the entries of the tensor $\mc O^{\TVect}_{\TVect_G}$ more explicitly, but it is not required for our study and we find the general form to be more revealing. In particular, we know that for an indecomposable $\Vect_G$-module category, the non-trivial hom-spaces appearing in eq.~\eqref{eq:prism} are all one-dimensional and as such we could have omitted to include labels for the nodes.
 
Crucially, the 2-category $\TRep(G)$---in contrast to $\TVect_G$---has non-trivial 1-morphisms, which have not played any role so far. Given a right $\Vect_G$-module 1-category $\mc N$, we have in particular
\begin{equation}
    \HomC_{\TRep(G)}(\mc N,\mc N) = \Fun_{\Vect_G}(\mc N, \mc N) =: (\Vect_G)^\star_{\mc N} \, ,
\end{equation}
i.e. the hom-category of endomorphisms of the simple object $\mc N$ is given by the Morita dual $\Vect_G$ w.r.t. $\mc N$ \cite{etingof2016tensor}.
It turns out that simple objects in $(\Vect_G)^\star_{\mc N}$ label \emph{higher-order} virtual symmetry operators. Indeed, $\mc O^{\TVect}_{\TVect_G}$ itself satisfies symmetry conditions w.r.t. one-dimensional virtual operators of the form
\begin{equation}
    \includeTikz{0}{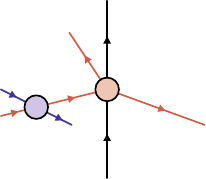}{\higherPullingThrough{1}} 
    \; \stackrel{!}{=} \;  
    \includeTikz{0}{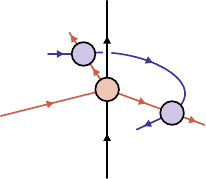}{\higherPullingThrough{2}} \, .
\end{equation}
Treating the tensor $\mc O^{\TVect}_{\TVect_G}$ as a two-dimensional tensor where the membranes $(\snum{01})$, $(\snum{12})$ and $(\snum{02})$ now correspond to physical indices, these higher-order virtual operators can be inferred from the study conducted in \cite{Lootens:2020mso}. First, notice that the right $\Vect_G$-module category $\mc N$ naturally possesses the structure of a left module 1-category over $(\Vect_G)^\star_\mc N$ so that $\mc N$ is also a $((\Vect_G)^\star_\mc N,\Vect_G)$-bimodule 1-category. The virtual symmetries of $\mc O^{\TVect}_{\TVect_G}$ are then given in terms of tensors labelled by simple objects $V \in \sOb((\Vect_G)^\star_\mc N)$ whose non-vanishing components evaluate to the matrix elements of maps specifying the bimodule associators $\alpha^{\caact}_{V,M,\mathbb C_g}$ of $\mc N$. Graphically, these non-vanishing components read\footnote{Interestingly, the graphical calculus we have been employing make apparent the fact that virtual operators of the tensor $\mc O^{\TVect}_{\TVect_G}$ are labelled by module endofunctors over $\Vect_G$. This can be used to confirm a posteriori the definition of the virtual operators in two-dimensional tensor network states.}
\begin{equation}
	\includeTikz{0}{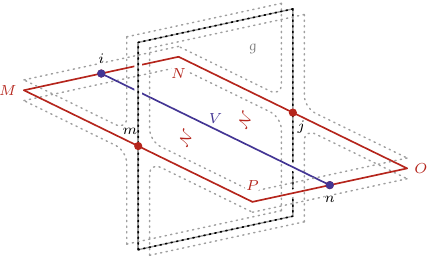}{\cube}= \SixJCaact{V}{M}{N}{\mathbb C_g}{O}{P}_{ij,mn} \, ,
\end{equation}
for any $g \in G$, simple object $V \in \sOb((\Vect_G)^\star_\mc N)$, simple objects $M,N,O,P \in {\rm sOb}(\mc N)$, and basis vectors $|{}^{VM}_N ,i\ra \in \Hom_{\mc N}(V\act M,N)$, 
$|{}^{N \mathbb C_g}_O ,j\ra \in \Hom_{\mc N}(N \cat \mathbb C_g ,O)$, 
$\la {}^{M \mathbb C_g}_P,m | \in \Hom_{\mc N}(M \cat \mathbb C_g,P)^\star$, 
$\la {}^{VP}_{O},n | \in \Hom_{\mc N}(V \act P,O)^\star$. The higher-order pulling-through condition is provided by the right analogue of eq.~\eqref{eq:SixJBiEq}, which is fulfilled in virtue of the pentagon axiom involving $\alpha^{\caact}$ and $\alpha^{\cat}$. For instance, the component of the tensor $\mc O^{\TVect}_{\TVect_G}$ labelled by $\mc N= \Vect$ satisfies symmetry conditions with respect to virtual operators labelled by simple objects $V$ in $(\Vect_G)^\star_{\Vect} \cong \Rep(G) \cong \Mod(\mathbb C[G])$. In this case, the matrix elements of the bimodule associator $\alpha^{\caact}$ boils down to matrix elements $\rho(g)_{i,j}$ satisfying $g \act | j \ra = \sum_{i=i}^{{\rm dim}_{\mathbb C} \, V} \rho(g)_{i,j} |i \ra$, for any $|i\ra \in V$, and the higher-order pulling-through condition simply amounts to the defining property of group representations, namely $\rho(gh)_{i,k} = \sum_{j=1}^{{\rm dim}_{\mathbb C} \, V}\rho(g)_{i,j} \rho(h)_{j,k}$.

Finally, we would like to consider the possibility of `fusing' two such virtual operators. Let us suppose that we  successively contract the tensor $\mc T^{\TVect}_{\TVect_G}$ with two copies of $\mc O^{\TVect}_{\TVect_G}$ along the same set of virtual indices.  We should be able to express the successive action of these virtual operators as that of a single virtual operator. This is indeed possible due to the fact that $\TRep(G)$---as a spherical fusion 2-category---can be equipped with the structure of a semi-simple monoidal 2-category such that the monoidal product is obtained by defining a $\Vect_G$-module structure on $\mc N \boxtimes \mc N'$ via $(N \boxtimes N') \cat \mathbb C_g := (N \cat \mathbb C_g) \boxtimes (N' \cat \mathbb C_g)$ for any $N \in \mc N$, $N' \in \mc N'$ and $\mc N,\mc N' \in \sOb(\TRep(G))$. The semi-simplicity of $\TRep(G)$ ensures that such a monoidal product of two simple objects can be decomposed into a direct sum of indecomposable $\Vect_G$-module categories. In terms of tensors, this fusion process is realised via a tensor whose non-vanishing components depicted as
\begin{equation}
	\includeTikz{0}{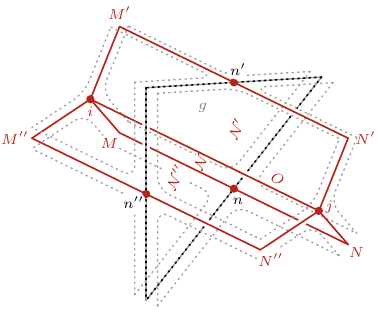}{\fusion}
\end{equation}
evaluate to the matrix entries of the  bimodule \emph{right} pentagonator components $\tilde \pi^{\caact}_{\Vect_g,\Vect, \mc N, \mc N'}$ (see def.~\ref{def:biMod2Cat}) for any simple object $\Vect_g \in \sOb(\TVect_G)$, indecomposable right $\Vect_G$-module categories $\mc N, \mc N', \mc N'' \in  \sOb(\TRep(G))$, $M,N \in \sOb(\mc N)$, simple objects $M',N' \in \sOb(\mc N')$, $M'',N'' \in \sOb(\mc N'')$, $O \in \sOb(\Fun_{\Vect_G}(\mc N' \boxtimes \mc N,\mc N''))$, basis vectors
$| {}^{M\mathbb C_g}_N,n  \ra \in \Hom_{\mc N}( M \cat \mathbb C_g ,N)$,
$| {}^{M'\mathbb C_g}_{N'},n'  \ra \in \Hom_{\mc N}( M' \cat \mathbb C_g ,N')$,
$\la {}^{M''\mathbb C_g}_{N''},n'' | \in \Hom_{\mc N''}( M'' \cat \mathbb C_g ,N'')^\star$, as well as basis vectors labelled by $i$ and $j$ in  the hom-spaces $\Hom_{\TRep(G)} ( (M' \cat {\rm id}_{\mc N}) \circ M, ({\rm id}_{\Vect} \cat O) \circ M'')$ and $\Hom_{\TRep(G)}(({\rm id}_{\Vect} \cat O) \circ N'', (N' \cat {\rm id}_{\mc N})\circ N)$, respectively. Interestingly, it follows from the associahedron axiom involving $\pi^{\caact}$ and $\tilde \pi^{\caact}$ that this fusion tensor satisfies intertwining conditions of the form
\begin{equation}
	\includeTikz{0}{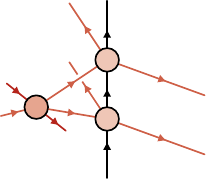}{\zipper{1}} 
	\; \stackrel{!}{=} \;  
	\includeTikz{0}{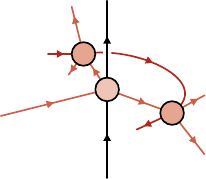}{\zipper{2}} \, ,
\end{equation}
where we introduced a more conventional notation (dark red blobs) for the fusion tensors. 
This fusion rule is associative and the fusion tensors satisfy pulling-through conditions of the form \eqref{eq:pullingThrough} w.r.t. tensors that evaluate to the components of the module pentagonator $\pi^{\cat}$ of $\TVect$  as a right module category over $\TRep(G)$.

\subsection{Morita equivalence}

Guided by the derivations of the previous section, we shall now explicitly establish the \emph{Morita equivalence} between the 2-categories $\TVect_G$ and $\TRep(G)$, that is the existence of a module 2-category $\mc M$ such that $(\TVect_G)^\star_\mc M := \msf{2Fun}_{\TVect_G}(\mc M,\mc M)$ is equivalent as a 2-category to $\TRep(G) = \MOD(\Vect_G^{\rm op})$. The demonstration is inspired by that of the Morita equivalence between the (1-)categories $\Vect_G$ and $\msf{Rep}(G) \cong \Mod(\mathbb C[G])$ \cite{etingof2016tensor}.

\medskip \noindent
Recall that via the  forgetful monoidal 2-functor $\TVect_G \to \TVect$, the 2-category $\TVect$ defines a (finite semi-simple) module 2-category over $\TVect_G$ with trivial $G$-action 2-functor such that the module associator and pentagonator evaluate to the identity 1- and 2-morphisms, respectively. By definition, a $\TVect_G$-module 2-endofunctor of $\TVect$ consists of a 2-functor $F:\TVect \to \TVect$ that is specified by the 2-vector space $\msf V :=F(\Vect)$, an adjoint 2-natural equivalence $\omega$ prescribed by
\begin{equation}
    \omega_g  \in \HomC_{\TVect}\! \big(F(\Vect_g \boxtimes \Vect), \Vect_g \boxtimes \, F(\Vect)\big) = \msf{Fun}_{\Vect}(\msf V, \msf V) \equiv \msf{Fun}(\msf V , \msf V)
     \,, \q \forall \, g \in G \, , 
\end{equation}
and an invertible modification $\Omega$ whose components are provided by natural transformations
\begin{equation}
    \Omega_{g,h} \in \Hom_{\msf{Fun}(\msf V, \msf V)} \big(\omega_g \circ \omega_h, \omega_{gh} \big) \, , \q \forall \, g,h \in G \, .
\end{equation}
It follows from the associahedron axiom in def.~\ref{def:mod2Cat2Fun} of  a module 2-category 2-functor that the natural transformations $\Omega_{g,h}$ satisfies the 2-cocycle condition $\Omega_{h,k} \circ \Omega_{g,hk} =  \Omega_{g,h} \circ  \Omega_{gh,k}$, as natural isomorphisms $\omega_{g} \circ \omega_{h} \circ \omega_h \xrightarrow{\sim} \omega_{ghk}$ for all $g,h,k \in G$.
Introducing the notations $\cat: \msf V \times \Vect_G  \to \msf V$, whereby $M \cat \mathbb C_g := \omega_g(M)$, and $\alpha^\cat_{M,\mathbb C_g, \mathbb C_h} := (\Omega_{g,h})_M$ the 2-cocycle condition above implies that the equality
\begin{equation}
	(\alpha^{\cat}_{M,\mathbb C_g,\mathbb C_h} \cat {\rm id}_{\mathbb C_k}) \circ \alpha^{\cat}_{M,\mathbb C_{gh},\mathbb C_k} \circ {\rm id}_{M \cat \mathbb C_{ghk}}
	=
	\alpha^{\cat}_{M \cat \mathbb C_g,\mathbb C_h,\mathbb C_k} \circ\alpha^{\cat}_{M,\mathbb C_g,\mathbb C_{hk}}
\end{equation}
holds in $\Hom_{\msf V}( ((M \cat \mathbb C_g)\cat \mathbb C_h)\cat \mathbb C_k, M \cat \mathbb C_{g(hk)} )$ for all $g,h,k \in G$ and $M \in \Ob(\msf V)$. It follows that the triple $(\msf V, \cat, \alpha^{\cat})$ thus constructed defines a \emph{right} module category over the monoidal category $\Vect_G$.\footnote{Recall that by definition the monoidal associator of $\Vect_G$ evaluates to the identity 1-morphism.} Conversely, any right $\Vect_G$-module category determines a $\TVect_G$-module 2-endofunctor of $\TVect$.

Given a pair $(F,\omega,\Omega)$ and $(F',\omega',\Omega')$ of $\TVect_G$-module 2-endofunctors of $\TVect$ identified with $\Vect_G$-module categories $(\msf V, \cat, \alpha^\cat)$ and $(\msf V',\catb, \alpha^{\catbsss})$, respectively, a $\TVect_G$-module 2-natural transformation between them consists of a 2-natural transformation $\theta: F \Rightarrow F'$, which is specified by a choice of (1-)functor $\widetilde F \in \msf{Fun}(\msf V, \msf V')$, as well as an invertible modification $\Theta$ prescribed by a collection of natural transformations
\begin{equation}
    \Theta_g \in \Hom_{\msf{Fun}(\msf V,\msf V')} (\omega_g \circ \widetilde F, \widetilde F \circ \omega'_g) \, , \q \forall \, g \in G \, .
\end{equation}
Moreover,  the horizontal composition $ \Omega \circ {\rm id}$ of the 2-morphism $\Omega$ with the trivial interchanger in the top diagram of the coherence axiom given in def.~\ref{def:mod2Cat2Nat} is specified by natural transformations $(\Omega \circ {\rm id})_{g,h}
\in \Hom_{\msf{Fun}(\msf V, \msf V')}(\omega_g \circ \omega_h \circ \widetilde F, \omega_{gh}\circ \widetilde F)$ so that $((\Omega \circ {\rm id})_{g,h})_M = \widetilde F(\alpha^{\cat}_{M,\mathbb C_g, \mathbb C_h})$. It follows from the defining axiom of the invertible modification $\Theta$ in def.~\ref{def:mod2Cat2Nat} of a module 2-category 2-natural transformation that $\Theta_h \circ \Theta_{g} \circ \Omega'_{g,h} = (\Omega \circ {\rm id})_{g,h} \circ \Theta_{gh}$ as natural isomorphisms $\omega_g \circ \omega_h \circ \widetilde F \xrightarrow{\sim} \widetilde F \circ \omega'_{gh}$, for all $g,h \in G$.
Introducing the notation $\widetilde \omega_{M,\mathbb C_g}:= (\Theta_g)_M$, the previous condition ensures that
\begin{equation}
	\widetilde \omega_{M \cat \mathbb C_g,\mathbb C_h} \circ (\widetilde \omega_{M,\mathbb C_g} \catb {\rm id}_{\mathbb C_h}) \circ \alpha^{\catbsss}_{\widetilde F(M),\mathbb C_g, \mathbb C_h}
	=
	\widetilde F(\alpha^{\cat}_{M,\mathbb C_g,\mathbb C_h}) \circ \widetilde \omega_{M,\mathbb C_{gh}}
\end{equation}
holds in $\Hom_{\msf V'}(\widetilde F((M \cat \mathbb C_g)\cat \mathbb C_h),\widetilde F(M)\cat \mathbb C_{gh})$
for all $g,h \in G$ and $M \in \Ob(\msf V)$. We thus conclude that the pair $(\widetilde F, \widetilde \omega)$ thus constructed defined a right $\Vect_G$-module  functor. Conversely, any right $\Vect_G$-module category functor determines a $\TVect_G$-module 2-natural transformation between $\TVect_G$-module 2-endofunctors of $\TVect$.

Finally, given a pair $(\theta, \Theta)$ and $(\theta', \Theta')$ of $\TVect_G$-module 2-natural transformations corresponding to right $\Vect_G$-module functors $(\widetilde F, \widetilde \omega)$ and $(\widehat F, \widehat \omega)$, a $\TVect_G$-module modification $\vartheta$ between them consists of a modification $\vartheta : \theta \Rrightarrow \theta'$, which is specified by a choice of natural transformation $\widetilde \theta \in \Hom_{\msf{Fun}(\msf V, \msf V')}(\widetilde F, \widehat F)$. It follows from the defining axiom of $\vartheta$ in def.~\ref{def:mod2Cat2Nat} of a module 2-category modification that
\begin{equation}
    \widetilde \theta_{M \cat \mathbb C_g} \circ \widehat \omega_{M , \mathbb C_g} = \widetilde \omega_{M,\mathbb C_g} \circ ( \widetilde \theta_M \catb{\rm id}_{\mathbb C_g})
\end{equation}
holds in $\Hom_{\msf{Fun}(\msf V, \msf V')}(\widetilde F(M \cat \mathbb C_g), \widehat F(M) \cat \mathbb C_g)$ for all $g \in G$ and $M \in \Ob(\msf V)$. Therefore, the natural transformation $\widetilde \theta$ is endowed with a right $\Vect_G$-module structure. Conversely, any right $\Vect_G$-module natural transformation determines a $\TVect_G$-module modification between $\TVect_G$-module 2-natural transformations of $\TVect_G$-module 2-endofunctors of $\TVect$. Putting everything together, we obtain the equivalence of 2-categories
\begin{equation}
    (\TVect_G)^\star_{\TVect} = \msf{2Fun}_{\TVect_G}(\TVect, \TVect) \cong \MOD(\Vect_G^{\rm op}) = \TRep(G) \, ,    
\end{equation}
from which the Morita equivalence between $\TVect_G$ and $\TRep(G)$ follows. Analogous---but much simpler---computations yields
\begin{equation}
    (\TVect_G)^\star_{\TVect_G} = \msf{2Fun}_{\TVect_G}(\TVect_G, \TVect_G) \cong \TVect_G^{\rm op} \, , 
\end{equation}
as expected. 

\medskip \noindent
Putting everything together, we constructed two canonical tensor network representations of the membrane-net state space $\mc M[\Sigma,\TVect_G]$ whose 2-categories of virtual operators, namely $\TVect_G$ and $\TRep(G)$, are Morita equivalent. Furthermore, it can be inferred from our derivations that these 2-categories correspond to those of boundary operators for the Dirichlet and Neumann boundary conditions of the topological model, respectively, so that the existence of these two tensor network representations is a manifestation of the electromagnetic duality of the system.

In light of these computations, let us make a few comments about the general case. Given an arbitrary left $\TVect_G$-module 2-category, we expect the virtual symmetry operators of the tensor network representation given by $\mc T^{\mc M}_{\TVect_G}$ to be encoded into the Morita dual of $\TVect_G$ with respect to $\mc M$, namely $(\TVect_G)^{\star}_{\mc M}$. By construction, $\mc M$ has the structure of a right module 2-category over $(\TVect_G)^\star_\mc M$ so that $\mc M$ is also a $(\Vect_G,(\TVect_G)^\star_\mc M)$-bimodule 2-category. The tensor components of $\mc O^{\mc M}_{\TVect_G}$ are then expected to correspond to matrix elements of maps specifying the bimodule left pentagonator $\pi^{\caact}$ so that the pulling-through condition is ensured by the associahedron axiom involving $\pi^{\caact}$ and $\pi^{\act}$. The fusion of virtual operators is then performed via a tensor whose components evaluate to the matrix elements of maps specifying the bimodule right pentagonator $\tilde \pi^{\caact}$.

\subsection{Three-dimensional toric code}

Let us now revisit the results obtained in \cite{Delcamp:2020rds} in light of the previous derivations. Given the (3+1)d toric code  Hamiltonian, the ground state subspace can be conveniently constructed by enforcing either the $X$ or $Z$ stabiliser constraints via a choice of product state in the relevant basis, and then project onto the subspace of the remaining stabiliser constraints. Writing the latter projector as a tensor network yields a tensor network representation of the ground state subspace. Depending on the stabiliser constraints that are initially imposed, we find two (canonical) tensor network representations characterised by distinct virtual symmetry conditions. The first representation satisfies a `global' virtual symmetry condition with respect to a membrane of $Z$ operators simultaneously acting on all the entanglement degrees of freedom, whereas the second representation satisfies a `local' virtual symmetry condition with respect to a string of $X$ operators acting on subsets of entanglement degrees of freedom.\footnote{Note that the choice of Pauli operators is simply a matter of convention, but we consistently use the same convention throughout.} Furthermore, non-annihilating operator insertions breaking these virtual symmetries lift to point-like electric or loop-like magnetic \emph{physical} excitations, respectively.

\bigskip \noindent
Let us now reveal the algebraic structures underlying these canonical tensor network representations by applying the previous program to the case $G = \mathbb Z_2$. The category theoretical input data of the (3+1)d toric code is the spherical fusion 2-category $\TVect_{\mathbb Z_2}$, such that the topological order is encoded into $\mc Z(\TVect_{\mathbb Z_2}) \cong \TRep(\mathbb Z_2) \boxplus \TRep(\mathbb Z_2)$, and the two canonical gapped boundary conditions are provided by the (finite semi-simple) module 2-categories $\TVect$ and $\TVect_{\mathbb Z_2}$, respectively. It was confirmed explicitly in \cite{kong2020defects} that the corresponding boundary operators and their statistics are encoded into the braided fusion 2-categories $\TRep(\mathbb Z_2)$ and $\TVect_{\mathbb Z_2}$, respectively. Let us now consider the tensor network representation $\mc T^{\TVect_{\mathbb Z_2}}_{\TVect_{\mathbb Z_2}}$ labelled by the regular module 2-category $\TVect_{\mathbb Z_2}$. Writing the group $\mathbb Z_2$ as $\{+,-\}$, it follows from the analysis carried out in the previous section that the component of the virtual symmetry tensor $\mc O_{\TVect_{\mathbb Z_2}}^{\TVect_{\mathbb Z_2}}$ labelled by the non-trivial simple object $\Vect_{-} \in \Ob(\TVect_{\mathbb Z_2})$ is effectively provided by a tensor product of Pauli $Z$ operators that act on the virtual membranes of the tensor $\mc T^{\TVect_{\mathbb Z_2}}_{\TVect_{\mathbb Z_2}}$. The corresponding pulling-through condition then yields the `global' symmetry condition mentioned above.

The second tensor network representation $\mc T_{\TVect_{\mathbb Z_2}}^{\TVect}$ is that labelled by the trivial module 2-category $\TVect$. We know from the analysis of the general case that the components of the virtual symmetry tensor $\mc O_{\TVect_{\mathbb Z_2}}^{\TVect}$ are labelled by simple objects in $(\TVect_{\mathbb Z_2})^\star_{\TVect} \cong \TRep(\mathbb Z_2)$, i.e. indecomposable right $\Vect_{\mathbb Z_2}$-module categories, and evaluate to the matrix entries of the corresponding  module associators. The 2-category $\TRep(\mathbb Z_2)$ has up to equivalence two simple objects, namely $\Vect$ and $\Vect_{\mathbb Z_2}$. Let focus on the virtual tensor associated with the simple object $\Vect \in \Ob(\TRep(\mathbb Z_2))$. The single simple object of $\Vect$ is trivially acted upon by the physical membranes of $\mc T^{\TVect}_{\TVect_{\mathbb Z_2}}$, whereas the tensor $\mc O^{\TVect}_{\TVect_{\mathbb Z_2}}$ evaluates to the matrix elements of the module associator of $\Vect$, which are all equal to 1, and thus the associated pulling-through condition is trivial. So how does one recover the expected virtual symmetry with respect to strings of $X$ operators? We need to consider the higher-order virtual symmetries described at the end of sec.~\ref{sec:virtual}. We established that these are labelled by simple endomorphisms of $\Vect$ in $\TRep(\mathbb Z_2)$, i.e. simple objects in
\begin{equation}
    \HomC_{\TRep(\mathbb Z_2)}(\Vect,\Vect) = \Fun_{\Vect_{\mathbb Z_2}}(\Vect, \Vect) = (\Vect_{\mathbb Z_2})^\star_{\Vect} \cong \Rep(\mathbb Z_2) \, .
\end{equation}
Writing $\mathbb Z_2 = \{+,-\}$ as before, the higher-order virtual operator labelled by the \emph{sign} representation of $\mathbb Z_2$ acts on the indices of $\mc O^{\TVect}_{\TVect_{\mathbb Z_2}}$ that are identified with the physical membranes of $\mc T^{\TVect}_{\TVect_{\mathbb Z_2}}$ upon contraction via Pauli $X$ operators. The corresponding pulling-through condition then amounts to the expected local symmetry condition with respect to closed one-dimensional loops of $X$ operators. Repeating the previous analysis for $\Vect_{\mathbb Z_2} \in \Ob(\TRep(\mathbb Z_2))$ does not yield any additional non-trivial virtual symmetry conditions for the tensor $\mc T_{\TVect_{\mathbb Z_2}}^{\TVect}$. This completes our analysis of this tensor network representation. 

Given a three-dimensional surface $\Sigma$, we mentioned in the introduction that different tensor network representations could be used over distinct submanifolds of $\Sigma$. This requires the introduction of virtual tensors that intertwine the different tensor network representations. In the case of the toric code---and more generally for the two canonical tensor network representations of topological gauge models we have been considering---these intertwining tensors admit a particularly simple form. In the case of the toric code, the non-vanishing components are of the form
\begin{equation}
    \includeTikz{0}{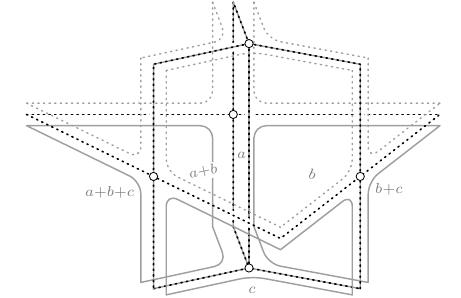}{\intertwiner} = 1 \, , \q \forall \, a,b,c \in \mathbb Z_2 \, .
\end{equation}
where we should think of the `horizontal' membrane as being labelled by the unique simple object in $\msf{2Fun}_{\TVect_{\mathbb Z_2}}(\TVect_{\mathbb Z_2},\TVect) \cong \TVect$ and the tensor as evaluating to the matrix entries of the invertible modification $\Omega$ of the corresponding $\TVect_{\mathbb Z_2}$-module 2-category 2-functor.
A network of such tensors intertwine the representation $\mc T^{\TVect}_{\TVect_{\mathbb Z_2}}$ (top) with the representation $\mc T^{\TVect_{\mathbb Z_2}}_{\TVect_{\mathbb Z_2}}$ (bottom). When interpreting this intertwining tensor network as a map from virtual to physical membranes acting on a two-dimensional spin model, it performs the duality from the (2+1)d Ising model to Wegner's $\mathbb Z_2$ gauge theory \cite{Wegner:1984qt,RevModPhys.51.659,Fisher2004}. A twisted version of this gauging map can be obtained by considering tensors of the form \eqref{eq:intertwiner} whose entries evaluate to $\alpha(a,b,c)$, where $\alpha$ is the non-trivial 3-cocycle in $H^3(\mathbb Z_2,{\rm U}(1))$. These intertwine the tensor network representations of the (3+1)d toric code associated with the $\TVect_{\mathbb Z_2}$-module categories $\TVect_{\mathbb Z_2}$ and $\TVect^\alpha$, respectively, as constructed in ex.~\ref{ex:TVecGModule}.

At this point, it is interesting to point out an important distinction between tensor network representations of two- and three-dimensional toric codes. The strategy spelt out above also yields two canonical tensor network representations of the (2+1)d toric code obtained by choosing the $\Vect_{\mathbb Z_2}$-module categories $\Vect$ and $\Vect_{\mathbb Z_2}$, respectively. However, both representations are characterised---up to an obvious change of basis---by the same virtual symmetry condition. This follows in particular from the fact that the categories of virtual operators $\Vect_{\mathbb Z_2}$ and $\Rep(\mathbb Z_2)$ are \emph{monoidally equivalent} in addition to being Morita equivalent.\footnote{More precisely, there is a monoidal equivalence between $\Rep(\mathbb Z_2)$ and $\Vect_{\mathbb Z_2^\vee}$ with $\mathbb Z_2^\vee = \Hom(\mathbb Z_2, \rU(1)) \simeq \mathbb Z_2$.} In terms of the quasi-one-dimensional system at the boundary, this monoidal equivalence is required for the self-duality of the (1+1)d Ising model. Conversely, the fact that the virtual operators in (3+1)d associated with each representation differ is because $\TVect_{\mathbb Z_2}$ and $\TRep(\mathbb Z_2)$ are not monoidally equivalent, but they are Morita equivalent, which ensures that they yield the same topological order. This failure to being monoidally equivalent in turn motivates the fact that the (2+1)d Ising model is not self-dual, but rather dual to Wegner's $\mathbb Z_2$-gauge theory as we just reviewed. 

Given the Morita equivalence between $\TVect_{\mathbb Z_2}$ and $\TRep(\mathbb Z_2)$, we may wish to consider the lattice model whose input spherical 2-category is $\TRep(\mathbb Z_2)$. However, since the virtual operator labelled by $\Vect_{\mathbb Z_2} \in \Ob(\TRep(\mathbb Z_2))$ does not yield any additional virtual symmetries, it is tempting to rather consider a sub-2-category of $\TRep(\mathbb Z_2)$. More precisely, we would like to exploit the fact that $\TRep(\mathbb Z_2) = \MOD(\Vect_{\mathbb Z_2}^{\rm op})$ can be thought as the \emph{idempotent completion} of a \emph{prefusion} 2-category, namely the \emph{delooping} $\msf{BVec}_{\mathbb Z_2}$ of $\Vect_{\mathbb Z_2}$ thought as a braided fusion category with the symmetric braiding \cite{douglas2018fusion, Johnson-Freyd:2020twl}.\footnote{Given a braided monoidal category $\mc C$, its delooping $\msf B \mc C$ is the monoidal bicategory consisting of a single object $\bul$ with $\HomC_{\msf B \mc C}(\bul) = \mc C$ such that the horizontal composition, the vertical composition and the monoidal product are provided by the monoidal product in $\mc C$, the composition of morphisms in $\mc C$ and the braiding structure in $\mc C$, respectively. The reason $\msf{BVec_{\mathbb Z_2}}$ is not a fusion 2-category is its failure to be semi-simple, which is remedied by the idempotent completion.} Crucially, it is expected that a fusion 2-category and its underlying prefusion 2-category yields the same four-dimensional state-sum invariant \cite{douglas2018fusion}. As such, instead of considering the Hamiltonian model with input data $\TRep(\mathbb Z_2)$, we can alternatively consider that with input data $\msf{BVec}_{\mathbb Z_2}$.  This model coincides with the \emph{Walker-Wang} model with input \emph{premodular category} $\Vect_{\mathbb Z_2}$ \cite{Walker:2011mda}, which is equivalent to a 2-form $\mathbb Z_2$-gauge model with trivial input (2-form) cocycle in $H^4(K(\mathbb Z_2,2),{\rm U}(1))$ \cite{Delcamp2019}. Given its 2-form gauge theory interpretation, this model hosts (bulk) loop-like charges and point-like fluxes. It is straightforward to confirm via Poincar\'e duality and $\mathbb Z_2$ Fourier transform that this model is dual to the (3+1)d toric code, as expected from the Morita equivalence between $\TRep(\mathbb Z_2)$ and $\TVect_{\mathbb Z_2}$. Let us conclude this section by writing down the tensor network representations of this $\mathbb Z_2$ 2-form model. As for the toric code, we distinguish two tensor representations with non-vanishing components given by
\begin{equation}
    \includeTikz{0}{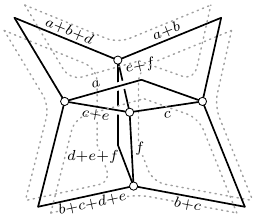}{\simplex{4}{}{}{}{}}
     \!\!\! = 1 \q {\rm and} \q 
     \includeTikz{0}{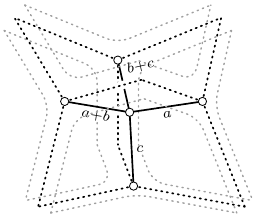}{\simplex{5}{}{}{}{}} \!\!\! = 1 
    \, ,
\end{equation}
for any $a,b,c,d,e,f \in \mathbb Z_2$, which satisfy virtual symmetry conditions with respect to string- and membrane-like operators, respectively.

\bigskip \noindent
\begin{center}
	\textbf{Acknowledgments}
\end{center}

\noindent
\emph{The author would like to thank Alex Bullivant, Thibault D\'ecoppet, Markus Hauru, Laurens Lootens, Norbert Schuch, Robijn Vanhove, Frank Verstraete and Dominic Williamson for fruitful discussions on closely related topics. He is particularly grateful to Laurens Lootens for insightful conversations about the two-dimensional case and for remarks on an earlier version of this manuscript, as well as Thibault D\'ecoppet for commenting on the role of separable algebras in the 2-categorical setting.
This research was partially funded by the Deutsche Forschungsgemeinschaft (DFG, German Research Foundation) under Germany’s Excellence Strategy – EXC-2111 – 390814868.}

\newpage

\titleformat{name=\section}[display]
{\normalfont}
{\footnotesize\centering {APPENDIX \thesection}}
{0pt}
{\large\bfseries\centering}

\newpage

\renewcommand*\refname{\hfill References \hfill}
\bibliographystyle{JHEP}
\bibliography{refs}

\end{document}